\documentclass[twocolumn]{IEEEtran}
 
\usepackage[T1]{fontenc}
\usepackage[latin1]{inputenc}
\usepackage{amsmath}
\usepackage{amsthm}
\usepackage{dsfont}
\usepackage{amsfonts}
\usepackage{mathrsfs}
\usepackage{pifont}
\usepackage{amssymb}
\usepackage{verbatim}
\usepackage{upgreek}
\usepackage[final]{graphicx}
\usepackage{subfigure}
\usepackage{epsfig}
\usepackage{booktabs}
\usepackage{bm}
\usepackage{float} 
\usepackage{color}
\usepackage{multirow}
\usepackage{fancyhdr}
\usepackage{cite}
\usepackage{balance}
\usepackage{algorithm}
\usepackage{algorithmic}
\usepackage[nolist,nohyperlinks]{acronym}
\usepackage[hyphens]{url}
\usepackage[hyperindex,breaklinks]{hyperref}
\hypersetup{breaklinks=true}
\usepackage{orcidlink}

\begin{acronym}
\acro{AWGN}{additive white Gaussian noise}
\acro{CDA}{continuous double auction}
\acro{CDAPS}{continuous double auction parameter selection}
\acro{COM}{communication}
\acro{ECCM}{electronic counter-counter measure}
\acro{EM}{electromagnetic}
\acro{ESA}{electronically steered array}
\acro{FFD}{far field distance}
\acro{ISAC}{integrated sensing and communication}
\acro{KKT}{Karush-Kuhn-Tucker}
\acro{LOS}{line of sight}
\acro{LPI}{low probability of intercept}
\acro{MIMO}{multiple input multiple output}
\acro{MPAR}{multifunction phased array radar}
\acro{NLOS}{non line of sight}
\acro{NP}{non-polynomial}
\acro{PAP}{power-aperture product}
\acro{$P_d$}{probability of detection}
\acro{PRI}{pulse repetition interval}
\acro{Q-RAM}{quality of service resource allocation method}
\acro{QoS}{quality-of-service} 
\acro{RCS}{radar cross section}
\acro{RF}{radio frequency}
\acro{RIS}{reflective intelligent surfaces}
\acro{RRE}{radar range equation}
\acro{RRM}{radar resource manager}
\acro{SIC}{successive interference cancellation}
\acro{SINR}{signal to interference plus noise ratio}
\acro{SNR}{signal to noise ratio}
\acro{SW}{Swerling}
\acro{UAV}{unmanned aerial vehicle}
\acro{URA}{uniform rectangular array}
 \end{acronym}

\newcommand{\ba}{\mbox{\boldmath{$a$}}}

\newcommand{\bs}{\mbox{\boldmath{$s$}}}

\newcommand{\bx}{\mbox{\boldmath{$x$}}}
\newcommand{\bw}{\mbox{\boldmath{$w$}}}

\newcommand{\bzeta}{\mbox{\boldmath{$\zeta$}}}

\def\E{{\mathbb E}}

\begin{document}

\title{Power-Aperture Resource Allocation for a MPAR with Communications Capabilities}

\author{Augusto Aubry\orcidlink{0000-0002-5353-0481}, \IEEEmembership{Senior Member, IEEE}, Antonio De Maio\orcidlink{0000-0001-8421-3318}, \IEEEmembership{Fellow, IEEE}, and Luca Pallotta\orcidlink{0000-0002-6918-0383}, \IEEEmembership{Senior Member, IEEE}
\thanks{A. Aubry and A. De Maio are with the Department of Electrical Engineering and Information Technology (DIETI), Universit\`{a} degli Studi di Napoli ``Federico II'', via Claudio 21, I-80125 Napoli, Italy. E-mail: \{augusto.aubry, ademaio\}@unina.it.}
\thanks{L. Pallotta is with School of Engineering, University of Basilicata, via dell'Ateneo Lucano 10, 85100 Potenza, Italy. E-mail: luca.pallotta@unibas.it.}}

\markboth{IEEE Transactions on Vehicular Technology}{Aubry et al}

\maketitle

\begin{abstract}
Multifunction phased array radars (MPARs) exploit the intrinsic flexibility of their active electronically steered array (ESA) to perform, at the same time, a multitude of operations, such as search, tracking, fire control, classification, and communications. This paper aims at addressing the MPAR resource allocation so as to satisfy the quality of service (QoS) demanded by both line of sight (LOS) and reflective intelligent surfaces (RIS)-aided non line of sight (NLOS) search operations along with communications tasks. To this end, the ranges at which the cumulative detection probability and the channel capacity per bandwidth reach a desired value are introduced as task quality metrics for the search and communication functions, respectively. Then, to quantify the satisfaction level of each task, for each of them a bespoke utility function is defined to map the associated quality metric into the corresponding perceived utility. Hence, assigning different priority weights to each task, the resource allocation problem, in terms of radar power aperture (PAP) specification, is formulated as a constrained optimization problem whose solution optimizes the global radar QoS. Several simulations are conducted in scenarios of practical interest to prove the effectiveness of the approach.
\end{abstract}


\section{Introduction}

Modern radar systems are becoming more and more sophisticated due to the stressing requirement of multifunctionality which can be defined as the capability of performing and managing a multitude of different operations. This is becoming of vital importance both in the modern battlefield scenario, that could comprise a plethora of different challenging requirements so as to account for possibly different threats, and in civilian applications (e.g., for a radar in urban environment attempting to detect drones both in \ac{LOS} and \ac{NLOS} as well as sending (possibly unidirectionally) communication signals to vehicular systems to convey potentially situational awareness information). Therefore, the \ac{MPAR} must perform different functions, such as search, tracking, fire control, classification, \ac{COM}, \ac{ECCM}, and also a multitude of tasks associated with each radar function \cite{moo2015adaptive}. To realize the aforementioned operations, the radar exploits the intrinsic flexibility provided by its active \ac{ESA} antenna, which allows to synthesize multiple diverse beams, as well as to steer them into specific directions with negligible delays and without angular continuity requirements. Moreover, on the transmit side different waveforms, \ac{PRI}, dwells, and energy values can be used. The management of the system degrees of freedom is demanded to the \ac{RRM}, which assigns priorities to the functions and to the tasks composing them. Additionally, it performs their dynamic scheduling together with the parameter selection and optimization \cite{friedman2018double}. Accordingly, the mentioned functions and tasks are generally accomplished dedicating (to each of them) specific amounts of the available radar resources, for instance multiplexing them over different time intervals and/or looking angles. It is also clear that, in assigning the resource to each function/task, the \ac{RRM} has to comply with physical and technical constraints, so as to appropriately handle the limited resource budget and the task induced performance constraints. In this respect, the \ac{RRM} must decide, on the basis of the assigned priorities, for the optimal controllable resource allocation in order to guarantee the necessary quality for the high priority tasks at the expense of the others. Needless to say, in the scheduling process, once the resources to manage are specified, a tailored figure of merit for each involved task as well as the associated utility function must be defined to realize an optimized distribution of the available radar degrees of freedom \cite{charlish2017cognitive}. Additionally, priorities are represented via scalar weights associated with each task. Then, the optimization problem for the resource sharing is set-up on the basis of the above quantities, where the objective function that describes the satisfaction for the overall success of the radar mission is maximized \cite{charlish2017cognitive}. In this respect, the \ac{RRM} can use different optimization tools to perform resource allocation. Among them, it is worth mentioning the \ac{Q-RAM} \cite{rajkumar1997resource} and the \ac{CDAPS} \cite{charlish2011agent, charlish2011autonomous}.

The \ac{Q-RAM} consists of few steps to handle a constrained optimization problem for discrete parameter selection. In a nutshell, starting from the situation where the resource for each task is zero, it performs an iterative subdivision of the degrees of freedom to each task in the order specified by the highest to the lowest marginal utility. Once the available resource is entirely allocated, the algorithm ends. Other interesting applications of the \ac{Q-RAM} within the framework of radar resource management can be found in \cite{kuo2003real, shih2003scheduling, shih2003template, hansen2004resource, hansen2006resource, ghosh2006integrated, hoffmann2014resource, moo2015adaptive}. Analogously, the \ac{CDAPS} models the tasks as agents, each of them having its own resource to utilize. Since, the total amount of resources for all tasks should not exceed a specific quantity, the problem is tackled through the application of a \ac{CDA} market algorithm \cite{friedman2018double}. Some other interesting uses of the \ac{CDAPS} related to the radar resource management problem can be found in \cite{charlish2012multi, charlish2017array, charlish2020development}. Other studies devoted to the optimization of the power allocation in a distributed \ac{MIMO} system performing both radar and communication functions have also been developed in the last years \cite{ahmed2019distributed, shi2020power, wu2022resource}. In particular, in \cite{ahmed2019distributed}, an optimization problem is formulated to reach as better as possible the desired performance in terms of target detection along with the desired data rate of the communication function. Moreover, in \cite{shi2020power}, the allocation paradigm is modified to boost the performance of the distributed \ac{MIMO} system in terms of its \ac{LPI}. Finally, in \cite{wu2022resource}, the above described resource allocation is expanded to the context of multi-target tracking.

Unlike the mentioned references, in this paper, a \ac{QoS} optimization is developed for a suitable allocation of the resources in a \ac{MPAR} system performing \ac{ISAC} activities, via a multitude of  functions and tasks ranging from surveillance in both \ac{LOS} and \ac{NLOS} environments to possibly unidirectional data transmission operations. Specifically, this paper is framed in the context of a single \ac{RF} \ac{ISAC} platform where the resources of a common phased array are shared among different concurring tasks in a smart way so as to avoid mutual interference. This is a different operational mode as compared with the \ac{ISAC} approaches developed in \cite{liu2022integrated, 9829746, chepuri2022integrated, 9847100, 10052711, 10054402}, where the waveforms and/or \ac{RIS} elements are controlled for communication/sensing centric or co-design paradigms considering the sum or linear combination of the \ac{SINR} for the different tasks as objective function. To this end, following the lead of \cite{charlish2017cognitive}, after defining parameters characterizing multiple search sectors, \ac{RIS}-aided search, as well as multiuser \ac{COM} tasks, their respective quality metrics and utility functions are introduced. Hence, the resulting resource allocation is formulated as a constrained optimization problem, where the \ac{PAP} is distributed to allow maximization of the overall \ac{QoS}. Notably, the formulated resource allocation problem is characterized by a non-convex objective function also only available in an implicit form. Hence the resulting optimization program can only be tackled via numerical methods. Several case studies of practical interest are analyzed to demonstrate the validity of the approach.

The paper is organized as follows. In Section \ref{sec_probl}, the \ac{MPAR} system is presented and the \ac{QoS} optimization problem is formulated considering the \ac{PAP} as degree of freedom. Then, in Section \ref{sec_resource_man} the quality metrics are defined for each task together with their respective utility functions. The problem is particularized and solved for some case studies of practical interest in Section \ref{sec_study_case}. Finally, some concluding remarks are given in Section \ref{sec_concl}.

\section*{Notations}

\begin{center}
\begin{tabular}{c|l}
\textbf{symbol} & \textbf{description}\\
\hline\\
$\ba$ & vectors (i.e., boldface)\\
$(\cdot)^T$ & transpose operator\\
$(\cdot)^\dag$ & conjugate transpose operator\\
${\mathbb{R}}^N$ & set of $N$-dimensional vectors of real numbers\\
$|\cdot|$ & modulus of a complex number\\
$\|\cdot\|$ & Frobenius matrix norm\\
$\E\left[\cdot\right]$ & statistical expectation\\
$f^{-1}(\cdot)$ & inverse of a function $f(\cdot)$
\end{tabular}
\end{center}

\section{Problem statement}\label{sec_probl}

\begin{figure}[ht!]
\begin{center}
\includegraphics[width=0.99\columnwidth]{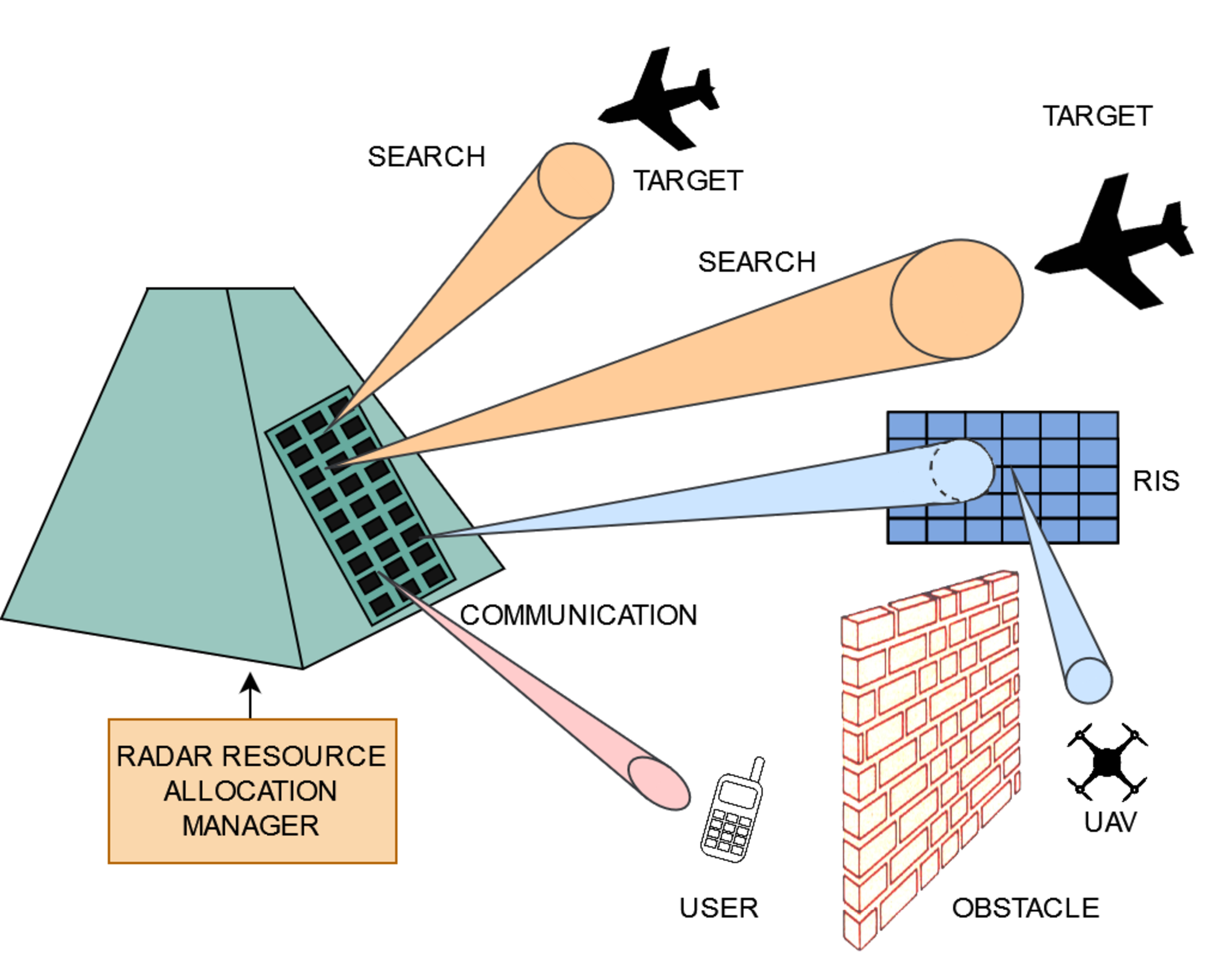}
\end{center}
\caption{A notional representation of a \ac{MPAR} system performing surveillance in \ac{LOS} situations, using \ac{RIS}s for \ac{NLOS} search, as well as implementing a \ac{COM} functionality.\label{schema_multifunction}}
\end{figure}

In this paper, a \ac{MPAR} system equipped with an active \ac{ESA} antenna is considered (see Fig. \ref{schema_multifunction} for a notional illustration of the operating scenario). It is capable of performing multiple functions, e.g., just to mention a few, radar surveillance (search) in \ac{LOS} scenarios, \ac{RIS}-aided search in \ac{NLOS} scenarios (a.k.a. detection \emph{over the corner}), \ac{COM} activities (possibly unidirectional) toward some users, tracking, and so on.

\begin{figure}[ht!]
\begin{center}
\includegraphics[width=0.85\columnwidth]{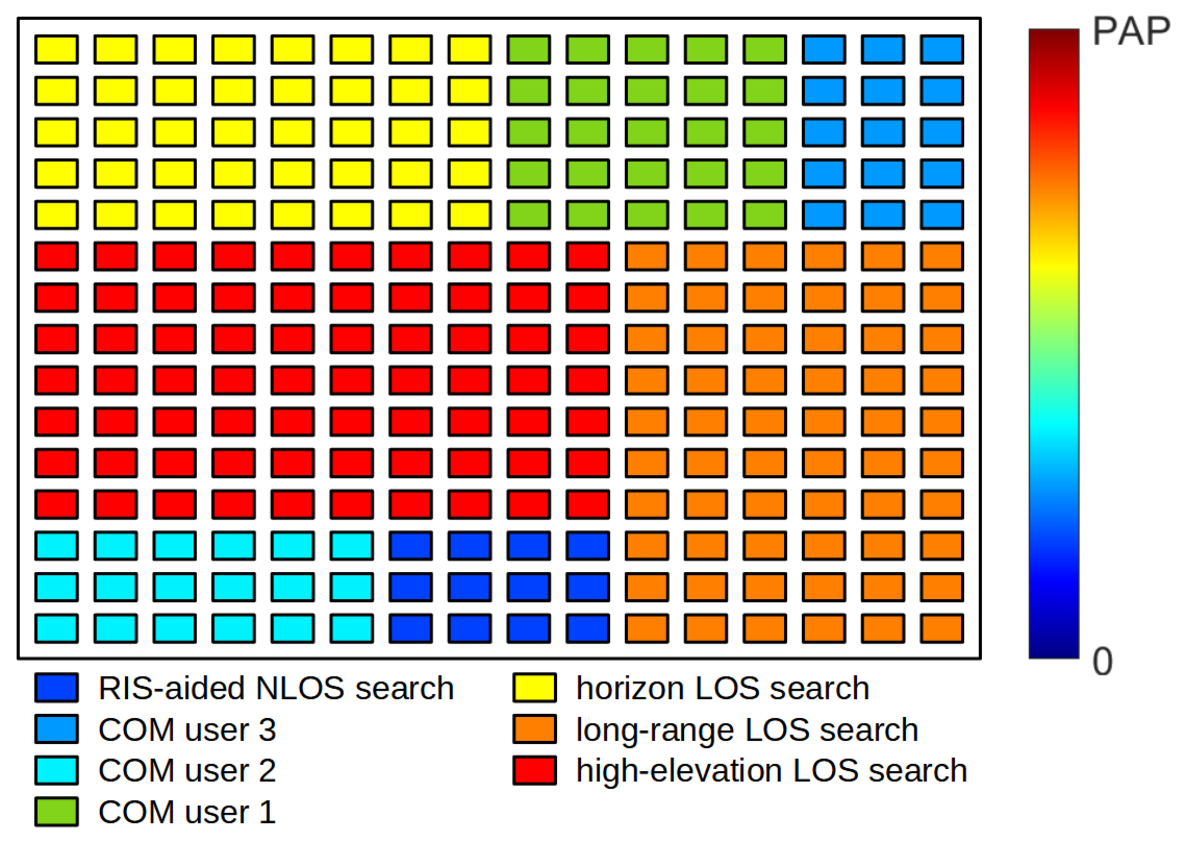}
\end{center}
\caption{A pictorial description of the \ac{PAP} allocation to the different tasks of the active \ac{ESA}.\label{allocazionePAP}}
\end{figure}

To allocate appropriately the resources required to each task, the radar employs a dynamic radar parameters assignment. In an ideal context, the system has the possibility to assign at each task the resources demanded to reach the desired performance. However, due to the limited availability, the radar system has to face with a suitable distribution of the degrees of freedom over the different tasks. Therefore in a \ac{MPAR}, the resources at the radar disposal are not a-priori fixed as in the classic surveillance systems, but rather they are dynamically allocated during its operation on the basis of the specific mission and its actual state, as well as depending on some priorities associated with each task. From a practical point of view the active \ac{ESA} is composed of many tiles each with a given \ac{PAP}. They are clustered according to the requirements of the system tasks so that each group realizes an overall \ac{PAP} value. A pictorial description of the concept can be seen in Fig. \ref{allocazionePAP}.

The \ac{PAP} (defined as the product between the average transmitted power and the radar aperture) is considered as the limited resource that must be granted to perform the different tasks. Obviously, if the available \ac{PAP} overcomes that needed to satisfy the requirements for all the active tasks, enough \ac{PAP} is given to each of them. Nevertheless, being the \ac{PAP} physically and practically limited, only a percentage of the resource demanded by each task can be, in general, allocated by the \ac{RRM} at each schedule time. The aforementioned assignment is performed on the basis of a pool of figure of merits and utility functions depending, in general, on the specific resources to distribute as well as on the design and environmental parameters (that are not under control), say $\bzeta_i$, $i=1,\ldots,L$, where $L$ is number of independent tasks, $T = \{T_1, \ldots, T_L\}$ that must share a finite common resource. To proceed further, recall that the $i$-th task utilizes the allocated resource to achieve a specific \ac{QoS}, quantified by a quality measure $q_i(\emph{\text{PAP}}_i;\bzeta_i)$ tailored to the specific task.
Therefore, the objective function for the resource allocation problem is obtained via the definition of a mapping among the $L$ task qualities and the achieved utilities in order to measure the overall effectiveness of the \ac{MPAR} mission. As a consequence, the \ac{RRM} should find the optimal partition of \ac{PAP} between tasks such that the weighted sum of their utilities is maximized \cite[Chap. 3]{charlish2017array}, \cite[Chap. 5]{charlish2017cognitive}. In this context, the task utility function provides the satisfaction level corresponding to the achieved task quality metric value. Moreover, to partially account for different degrees of relevance and priorities among the tasks, these utilities are suitably weighted in the formation of the overall \ac{RRM} utility metric. In other words, denoting by $\emph{\textbf{\text{PAP}}} = [\emph{\text{PAP}}_1,\emph{\text{PAP}}_2,\ldots, \emph{\text{PAP}}_L]^T \in \mathbb{R}^L$ the vector containing as $i$-th entry the \ac{PAP} attributed to the $i$-th task, $i=1,\ldots,L$, the \ac{PAP} distribution is obtained as the optimal solution to the following constrained optimization problem \cite[Chap. 5]{charlish2017cognitive}

\begin{equation}\label{eq_optimizationproblem}
\left\{
\begin{array}{ll}
& \displaystyle{\max_{\emph{\textbf{\text{PAP}}}} \, u(\emph{\textbf{\text{PAP}}})} \\
\text{s.t.} & \displaystyle{\sum_{i=1}^L \emph{\text{PAP}}_i \leq \emph{\text{PAP}}_{\emph{\text{tot}}}}\\
& \emph{\text{PAP}}_i \geq \emph{\text{PAP}}_{\text{min}_i}, i = 1, \ldots, L
\end{array}
\right.,
\end{equation}
where

\begin{equation}\label{eq_utility_total}
u(\emph{\textbf{\text{PAP}}}) = \displaystyle{\sum_{i=1}^L w_i u_i(q_i(\emph{\text{PAP}}_i; \bzeta_i))}
\end{equation}
$\emph{\text{PAP}}_{\emph{\text{tot}}}$ is the total amount of \ac{PAP} available at the \ac{MPAR}, $u_i(\cdot)$, $i=1,\ldots, L$, is the utility function of the $i$-th task, whereas $w_i$, $i=1,\ldots, L$, are the weights reflecting the priorities among the $L$ tasks. Finally, $\emph{\text{PAP}}_{\text{min}_i}$, $i = 1, \ldots, L$, guarantees that the $i$-th task is accomplished with a minimum level of \ac{QoS}. Note that, it is assumed $\sum_{i=1}^L \emph{\text{PAP}}_{\text{min}_i} \leq \emph{\text{PAP}}_{\emph{\text{tot}}}$, in order to ensure feasibility to the resource allocation problem.

Now, if the task utility function $u_i(q_i(\emph{\text{PAP}}_i; \bzeta_i))$ is a continuous convex function of $\emph{\text{PAP}}_i$, then the objective function is convex and hence the \ac{KKT} conditions can be exploited to establish the optimal resource allocation \cite[Chap. 5]{charlish2017cognitive}. If the resource, quality and utility functions are available in a closed-form, then the \ac{KKT} conditions can be solved analytically. However, it is often the case that the quality metrics do not possess a closed-form. In such a situation, even if the utilities exhibit a closed-form and the constraints are linear, the objective function is only available numerically, making the problem unsolvable in analytic form. As a consequence, the solution to the resource allocation problem can be only numerically obtained, as it is the case of the resource planning handled in this paper.

The next section describes the task quality metrics together with their corresponding utilities herein considered for the dynamic \ac{PAP} allocation paradigm described by \eqref{eq_optimizationproblem}.

\section{Task quality and utility for QoS resource management}\label{sec_resource_man}

The allocation strategy formalized by Problem \eqref{eq_optimizationproblem} depends on the considered figure of merits $q_i(\cdot;\cdot)$, and utility functions $u_i(\cdot)$, $i=1,\ldots, L$. The goal of this section is to specify them, so as to concretely define the scheduling machinery.

A meaningful figure of merit for the surveillance functions (both in the \ac{LOS} and \ac{NLOS} scenarios) is provided by the cumulative detection range, denoted as $R$, that is the range where the cumulative \ac{$P_d$} is larger than or equal to a desired value \cite{mallett1963cumulative, hoffmann2014resource, charlish2017array, charlish2017cognitive}. The cumulative \ac{$P_d$} is indeed defined as the probability that a target is detected at least once in a given number of dwells \cite{mallett1963cumulative, charlish2017cognitive}. In fact, when a target enters in a search sector, its detection can be performed over multiple scans. Moreover, the cumulative \ac{$P_d$} increases at each scan especially as the target approaches the radar.

Similarly, for the \ac{COM} function, the quality metric can be defined as the communication range, indicated as $R_{\text{com}}$, corresponding to the maximum distance at which a minimum bit-rate can be conveyed reliably. These two metrics are deeply described in Subsections \ref{secSearchTaskQualityMetric} and \ref{secCOMTaskQualityMetric}.

Before proceeding further, it is worth recalling the one-way link equation, which is useful for subsequent derivations.

\begin{figure}[ht!]
\begin{center}
\includegraphics[width=0.9\columnwidth]{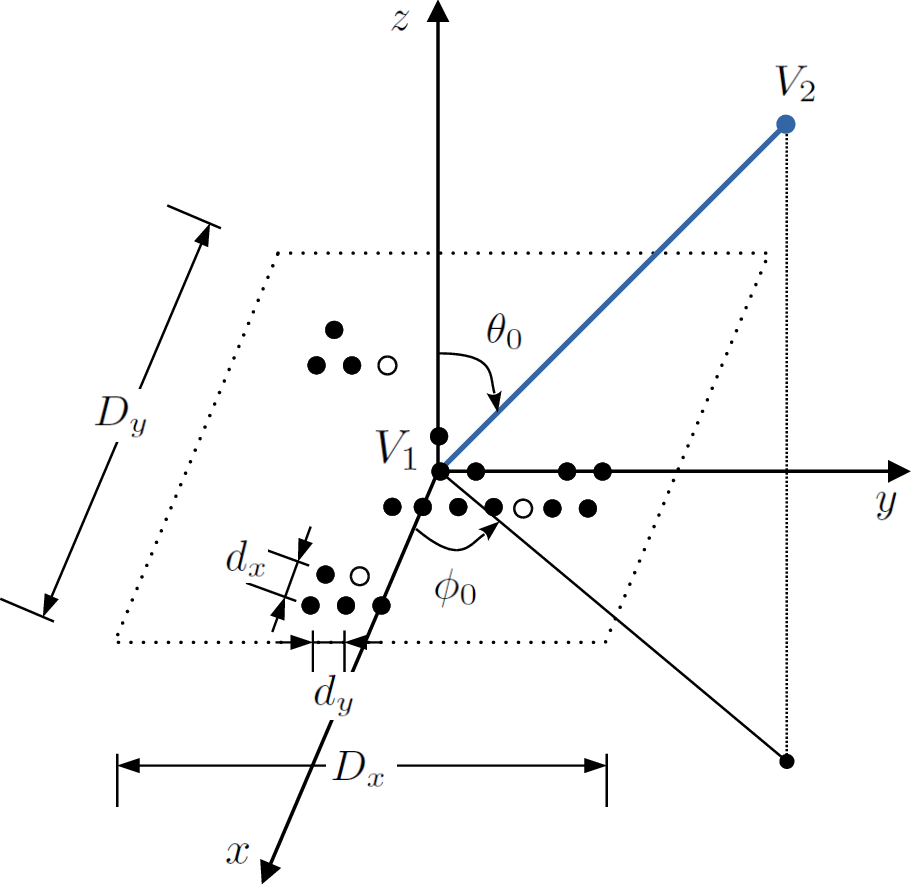}
\end{center}
\caption{Reference system of a generic planar array.\label{schema_array}}
\end{figure}

\textbf{Remark 1:} Let us consider a source located at the point $V_1 \in \mathbb{R}^3$, transmitting an \ac{EM} wave with a peak power of $P_T$ and an antenna steered in the direction described by the azimuth and elevation angles $\phi_0$ and $\theta_0$, according to the coordinate system depicted in Fig. \ref{schema_array}. Denoting by $G_T$ the peak antenna gain when it points in the boresight direction, the spatial power density at point $V_2$ is

\begin{equation}\label{eq_Pin}
\mathcal{P}^{in} = \frac{P_T G_T}{4\pi R^2 L_s L_{\text{steer}}},
\end{equation}
where $R=\|V_2-V_1\|$, i.e., the distance between the transmitter and the receiver, and $L_s$ is the combined system operational loss \cite{richards2010principles}. Moreover, $L_{\text{steer}}$ is the term accounting for the scanning gain loss of the steered antenna in the pointing direction\footnote{It is worth to underline that, even if $L_{\text{steer}}$ depends on the considered pointing angles, to simplify the notation, the dependence on $(\theta_0, \phi_0)$ is omitted in the rest of the paper.} $(\theta_0, \phi_0)$, which implicitly embeds the spatial selectivity in the antenna gain. In fact, as the pointing angles deviates from the boresight, the beam broadens while its peak drops out. The loss in peak gain due to scanning for a generic planar array depends on both the pointing direction (i.e., azimuth and elevation) and the single element radiating pattern. Practically, the values of these losses are off-line evaluated and then stored in a look-up table to be applied during radar's operation. However, in the particular case of a \ac{URA} under some technical assumptions as for instance large array size and omnidirectional array elements, $L_{\text{steer}}$ assumes a simplified approximated form, depending only on the elevation angle cosine \cite{elliott1963beamwidth, van2002optimum}.

\subsection{Search task quality metric}\label{secSearchTaskQualityMetric}

Let us indicate with $P_d(R')$ the single-look \ac{$P_d$} at range $R'$, and assume that $S$ is the number of scans the target needs to reach the range $R$ from the pop-up range $R_m$. Hence, the respective cumulative \ac{$P_d$} for the search sector of interest at range $R$ is given by \cite{mallett1963cumulative}

\begin{equation}\label{eq_cumulative}
P_c(R|R_m) = 1 - \prod_{n=0}^{S-1} \left[1 - P_d\left(R_m - n v_r t_f - \Delta \right)\right],
\end{equation}
with $v_r$ the target radial speed, $t_f$ the frame time (i.e., the time necessary to perform a single scan of the sector), and $\Delta$ a sample of a uniform random variable\footnote{Without loss of generality, $\Delta$ is set equal to zero in the next analyses.} in the interval $[0, v_r t_f]$, with $v_r t_f$ the distance traveled by the target in a single scan, modelling the initial target position in the corresponding radar cell. Note that, the functional dependence of the range on the pop-up range is $R=R_m-(S-1)v_r t_f - \Delta$. The single-look \ac{$P_d$} can be evaluated once the desired false alarm probability, say $P_{fa}$, is set. More specifically, assuming a \ac{SW} 0 (respectively a \ac{SW} 1) model for the target amplitude and assuming a coherent integration of the pulses in a dwell, the single-look detection probability at range $R'$ can be obtained as \cite{richards2010principles}

\begin{equation}\label{eq_SW0}
P_d(R') = Q_M\left(\sqrt{2\text{SNR}}, \sqrt{-2\log P_{fa}}\right)\quad (\text{SW}0)
\end{equation}
and 
\begin{equation}\label{eq_SW1}
P_d(R') = P_{fa}^{1/\left(1+ \text{SNR}\right)}\quad (\text{SW}1),
\end{equation}
where $Q_M{(\cdot, \cdot)}$ is the Marcum Q-function \cite{marcum1960statistical}. Note that, the functional dependence on the variable $R'$ of the $P_d$ is embedded in the expression of the coherent \ac{SNR}.

Let us now consider a radar located at point $V_1$ aimed at detecting a (possible) target at point $V_2$ in a \ac{LOS} environment. To contextualize the cumulative $P_d$ expression to the resource allocation process, it is necessary to particularize the result of \textbf{Remark 1} \eqref{eq_Pin} to the links $V_1$-$V_2$ and $V_2$-$V_1$. Accordingly, the \ac{SNR} can be expressed as \cite[eq. 2.17]{richards2010principles}

\begin{equation}\label{eq_SNR_radar}
\text{SNR}^{\text{LOS}} =  \frac{P_T G_T G_R \lambda_0^2 \sigma n_p}{(4\pi)^3 {R}_{\text{LOS}}^4 k_B T_s B L_s^{\text{LOS}}L_{\text{steer}}^{\text{LOS}}},
\end{equation}
where $G_R$ is the receiving antenna peak gain, ${R}_{\text{LOS}}=\|V_1 - V_2\|$, $T_s$ is the system noise temperature, $L_s^{\text{LOS}}$ is the combined two-way system operational loss \cite{richards2010principles}, $L_{\text{steer}}^{\text{LOS}}$ is the total scanning loss in the \ac{LOS} scenario, $\sigma$ is the target \ac{RCS}, $k_B$ is the Boltzmann's constant, $\lambda_0$ is the operating wavelength, and $n_p$ is the number of integrated pulses in a dwell. 
Assuming a monostatic radar configuration using the same beam in transmission and reception, \eqref{eq_SNR_radar} can be arranged in the search-form of the \ac{RRE} \cite{richards2010principles}. To this end, recall that \cite{richards2010principles}

\begin{equation}\label{eq_Td}
t_d = \frac{t_f}{M} = \frac{t_f}{\Omega A_e} \lambda_0^2,
\end{equation}
where $M$ is the number of beam positions to cover the solid angle search sector $\Omega$ and the effective area of the radar antenna $A_e$ is related to the radar peak gain by \cite{richards2010principles}

\begin{equation}\label{eq_gain}
G_T = 4\pi \frac{A_e}{\lambda_0^2}.
\end{equation}

Hence, substituting \eqref{eq_Td}-\eqref{eq_gain} in \eqref{eq_SNR_radar}, the search-form of the \ac{RRE} \eqref{eq_SW0}-\eqref{eq_SW1} boils down to

\begin{equation}\label{eq_SNR_radar_search}
\text{SNR}^{\text{LOS}} = \emph{\text{PAP}} \frac{\sigma}{4\pi k_B T_s {R}_{\text{LOS}}^4 L_s^{\text{LOS}}L_{\text{steer}}^{\text{LOS}} }\frac{t_f}{\Omega}
\end{equation}
where $\emph{\text{PAP}} = P_{avg} A_e$, with $P_{avg}$ the average transmit power. Before concluding the description of the \ac{LOS} scenario, it is worth to underline that \eqref{eq_SNR_radar_search} implicitly assumes the absence of interference among the signals of the different concurring tasks. In fact, the system manager itself coordinates the entire pool of sub-systems and allocates the resources of its phased array to avoid mutual interference among the different spatial beams.

As to the \ac{NLOS} scenario, encompassing a gapfiller \ac{RIS} that aids the detection \emph{over the corner} \cite{aubry2021reconfigurable}, let us indicate with $V_1$, $V_2$, and $V_3$ the positions of the radar, \ac{RIS}, and target, respectively, and, accordingly, $r_{\text{NLOS}} = \|V_1 - V_2\|$ and $R_{\text{NLOS}} = \|V_2 - V_3\|$. Therefore, leveraging \textbf{Remark 1}, the expression for the \ac{SNR} can be derived accounting for the multiple paths involved in the surveillance process, i.e., $V_1$-$V_2$, $V_2$-$V_3$, $V_3$-$V_2$, and $V_2$-$V_1$, along with the target \ac{RCS} and the radiation patterns synthesized at the \ac{RIS} equipment.\footnote{It is assumed that a \ac{RIS} realizes an appropriate beamforming, i.e., the parameter-settings of the \ac{RIS}, such as its element phase shifts, are already suitably optimized to face with the assigned task. In this respect, some techniques for \ac{RIS} phase-shift optimization can be exploited. The interested readers could refer to \cite{9500202, 9314027, 9926337, 10002901}, just to list a few.}
Specifically, the \ac{RRE} assumes the form \cite{aubry2021reconfigurable}

\begin{equation}\label{eq_SNR_RIS}
\text{SNR}^{\text{NLOS}} = \frac{G_T^2 G_{\text{RIS}}^2 A_{\text{RIS}}^2 \eta_{\text{RIS}}^2 \lambda_0^2 \sigma P_{avg} t_d}{r^4_{\text{NLOS}} R^4_{\text{NLOS}} (4\pi)^5 k_B T_s L_s^{\text{NLOS}}L_{\text{steer}}^{\text{NLOS}} },
\end{equation}
with $L_s^{\text{NLOS}}$ the combined system operational loss in the \ac{NLOS} case \cite{richards2010principles}, $L_{\text{steer}}^{\text{NLOS}}$ the total scanning loss in the \ac{NLOS} scenario. $A_{\text{RIS}}$ is the \ac{RIS} area, that for a uniform rectangular geometry can be expressed as $\delta_x \delta_y N_1 N_2$, with $\delta_x=\delta_y=\lambda_0/2$ the patch size along $x$- and $y$-direction, respectively, and $N_1$, $N_2$ the respective number of patches. Additionally, $\eta_{\text{RIS}}$ is the \ac{RIS} efficiency (assumed, for simplicity, common to all the patches), which accounts for taper and spillover effects \cite{chou2020aperture}. Hence, the product $A_{\text{RIS}} \eta_{\text{RIS}}$ is the effective aperture of the \ac{RIS}. Finally, $G_{\text{RIS}}$ is the \ac{RIS} peak gain.

The \ac{SNR} of a \ac{RIS}-aided search radar can be again expressed in terms of \ac{PAP}. Precisely, substituting \eqref{eq_Td}-\eqref{eq_gain} in \eqref{eq_SNR_RIS}, the search-form of the \ac{RIS}-aided \ac{RRE} is

\begin{equation}\label{eq_SNR_RIS_search}
\text{SNR}^{\text{NLOS}} = \frac{\emph{\text{PAP}}\, G_{\text{RIS}}^2 A_{\text{RIS}}^2 \eta_{\text{RIS}}^2 \sigma}{r^4_{\text{NLOS}} R^4_{\text{NLOS}} (4\pi)^3 k_B T_s L_s^{\text{NLOS}}
{L_{\text{steer}}^{\text{NLOS}}}} \frac{t_f}{\Omega}.
\end{equation}

Before concluding this section, it is now worth observing that a commonly reference value for the objective $P_c$ is 0.9. For this reason, the corresponding cumulative detection range denoted by $R_{90}^{\text{LOS}}$ for \ac{LOS} tasks, can be expressed as

\begin{equation}\label{eq_SNR_RIS_search1}
R_{90}^{\text{LOS}} = P_{c,{\text{LOS}}}^{-1}(0.9, R_m),
\end{equation}
having denoted by $P_{c,{\text{LOS}}}^{-1}(x|R_m)$ the inverse of the function in \eqref{eq_cumulative} for the \ac{LOS} case, i.e., when the SNR is dictated by $\text{SNR}^{\text{LOS}}$ in \eqref{eq_SNR_radar_search}. Analogously, for the \ac{NLOS} search task 

\begin{equation}\label{eq_SNR_RIS_search2}
{R}_{90}^{\text{NLOS}} = P_{c,{\text{NLOS}}}^{-1}(0.9, R_m),
\end{equation}
with $P_{c,{\text{NLOS}}}^{-1}(x|R_m)$ the inverse of the function in \eqref{eq_cumulative} for the \ac{NLOS} case, i.e., when the SNR is given by $\text{SNR}^{\text{NLOS}}$ in \eqref{eq_SNR_RIS_search}.

Note that equations \eqref{eq_SNR_RIS_search1} and \eqref{eq_SNR_RIS_search2} implicitly define the \ac{PAP}s demanded to attain the desired \ac{QoS}s for the surveillance tasks.

\subsection{COM task quality metric}\label{secCOMTaskQualityMetric}

The metric that describes the quality for a \ac{COM} task is the maximum range, indicated as $R_{\text{com}}$, for which the channel capacity per bandwidth is equal to a specific value. Before evaluating $R_{\text{com}}$, let us consider the transmission of a signal composed by the superposition of $U\leq B^{\text{COM}}T^{\text{sym}}$ frequency (or code) orthogonal waveforms, $x_i(t)$, $i=1,\ldots, U$, to $U$ \ac{COM} receiving users, with $B^{\text{COM}}$ the bandwidth reserved by the radar to \ac{COM} operations, and $T^{\text{sym}}$ the symbol interval. Then, the transmitted signal is 

\begin{equation}
\begin{split}
x(t) = \sum_{i=1}^U &\sum_{h=0}^{N^{\text{sym}}-1} s_i(\phi_i,\theta_i) x_i(t - h T^{\text{sym}})\alpha_i(h),\\
& 0\leq t \leq T^{\text{COM}},
\end{split}
\end{equation}
where $T^{\text{COM}} = T^{\text{sym}}N^{\text{COM}}$, $N^{\text{COM}}$ indicates the number of symbols transmitted in each scheduled interval, $\alpha_i(h)$, $h=0, \ldots, N^{\text{sym}}-1$, accounts for the information symbols for the $i$-th user, and $s_i(\theta_i, \phi_i)$ is the transmit beamformer pointing toward the $i$-th user at position $(\theta_i, \phi_i)$ w.r.t. the coordinate system centered at the transmitting antenna phase-center position.

Assuming an \ac{AWGN} channel, with $w(t)$ the noise contribution, the signal acquired at the $k$-th receiver, with reference to the $h$-th symbol interval, can be expressed as

\begin{equation}
\begin{split}
r_k(t) &= \bs^{\dag}_k\beta_k \bx(t-\tau_k) + w(t) \\
&= \beta_k \sum_{i=1}^U \bs^{\dag}_k\bs_i x_i(t-\tau_k)\alpha_i(h) + w(t),
\end{split}
\end{equation}
with $\bs_k$ the steering vector in the direction $(\theta_k,\phi_k)$, $\beta_k$ the complex scaling factor accounting for channel propagation effects and receive antenna, and $\tau_k$ the propagation time of the $k$-th user. Note that, the functional dependence of $\bs_k$ on $(\phi_k,\theta_k)$ is omitted for brevity.

At receiver side, the samples of the incoming signal after matched filter operation to $x_k(t-\tau_k)$ becomes

\begin{equation}
\langle r_k(t), x_k(t-\tau_k) \rangle = \beta_k g_k \alpha_k(h) + w_k(h),\; h = 0, \ldots, N^{\text{sym}}-1,
\end{equation} 
where $g_k$ is the transmitter beamformer complex gain in the direction $(\phi_k,\theta_k)$ of the $k$-th user, and $\langle \cdot, \cdot \rangle$ denotes the inner product operator.

Finally, the channel capacity per bandwidth (expressed in bit/s/Hz) for the $k$-th user can be defined as \cite{viswanath2002opportunistic, tse2005fundamentals, kountouris2008multiuser}

\begin{equation}\label{eq_channelCap}
C = \log_2\left(1 + \text{SNR}_k^{\text{COM}}\right),
\end{equation}
where $\text{SNR}_k^{\text{COM}}$ is the \ac{SNR} at the $k$-th \ac{COM} user receiver. 

Let us indicate with $V_1$ and $V_2$ the positions of the transmitter and the $k$-th \ac{COM} user, respectively, and $R_{k,{\text{COM}}} = \|V_1 - V_2\|$. According to \textbf{Remark 1}, the \ac{SNR} in \eqref{eq_channelCap} can be computed with respect to the link $V_1$-$V_2$ as

\begin{equation}\label{eq_sinr_com}
\text{SNR}_k^{\text{COM}}= \frac{P_k{\left|g_k\right|}^2 \left|\beta_k\right|^2}{\displaystyle{\sigma_k^2}},
\end{equation}
where $P_k = \E\left[\left|\alpha_k \right|^2\right]$ is the transmitting power for the $k$-th communication link, and $\sigma_k^2 = k_B T_s^{\text{COM}} B^{\text{COM}}$ is the noise power at the $k$-th receiver, with $T_s^{\text{COM}}$ and $B^{\text{COM}}$ the respective noise system temperature and effective bandwidth. Let us observe now that

$$
{\left|g_k\right|}^2 \left|\beta_k\right|^2 = \frac{G_T A^{\text{rx},k}_e}{4\pi R_{k,{\text{COM}}}^2 L_s^{\text{COM}}L_{\text{steer}}^{\text{COM}}}
$$
with $A^{\text{rx},k}_e$ the effective area of the $k$-th user receiving antenna, $L_s^{\text{COM}}$ the \ac{COM} system operational loss, and $L_{\text{steer}}^{\text{COM}}$ the total scanning loss in the \ac{COM} scenario. Hence, following the above definitions, $\text{SNR}_k$ in \eqref{eq_sinr_com} can be expressed in terms of \ac{PAP}, i.e.,

\begin{equation}
\text{SNR}_k = \frac{P_k {\left|g_k\right|}^2 \left|\beta_k\right|^2}{\displaystyle{\sigma_k^2}} = \emph{\text{PAP}}_k \frac{A^{\text{rx},k}_e}{\lambda_0^2 R_{k,{\text{COM}}}^2 L_s^{\text{COM}} L_{\text{steer}}^{\text{COM}} {\displaystyle{\sigma_k^2}}}.
\end{equation}

Finally, denoting by $C_{\text{desired}}$ the reference value for the objective channel capacity, its corresponding range, say $R_{\text{com}}$, is derived as follows

\begin{equation}
R_{\text{com}} = \sqrt{\frac{\emph{\text{PAP}}_k A^{\text{rx},k}_e}{\lambda_0^2 L_s^{\text{COM}} L_{\text{steer}}^{\text{COM}} \left(2^{C_{\text{desired}}}-1\right){\displaystyle{\sigma_k^2}}}}.
\end{equation}

\subsection{Task utility}\label{sec_utility}

Once the task quality metrics are defined, the joint optimum allocation of tasks' \ac{PAP}s can be computed as the optimal solution to the \ac{QoS} optimization problem in \eqref{eq_optimizationproblem}. In this respect, the \ac{RRM} needs to map the quality metrics to their corresponding utilities. As a matter of fact, the utility provides a description of the degree of satisfaction reached when each task is completed. A possible way to define the utility for the $i$-th considered task is through the following model \cite{hoffmann2014resource}

\begin{equation}\label{eq_utility_func}
\begin{split}
u_i(q_i(\text{PAP}_i, \bzeta_i)) &= u_i(R_{c}) \\
&= \left\{
\begin{matrix}
0, & R_{c} < R_{t_i}\\
\frac{R_{c} - R_{t_i}}{R_{o_i} - R_{t_i}}, & R_{t_i} \leq R_{c} \leq R_{o_i}\\
1, & R_{c} > R_{o_i}
\end{matrix}
\right.
\end{split}
\end{equation}
where $R_{t_i}$ and $R_{o_i}$ are the threshold and objective ranges of the $i$-th task, respectively. Moreover, $R_{c}$ denotes the quality metric for the specific task\footnote{Note that, the dependence on $\bzeta_i$ is omitted, being the environmental parameters fixed in the addressed problem.}, viz. the cumulative detection range $R_{\text{90}}$ or the communication range $R_{\text{com}}$, respectively. Obviously, at ranges lower than the threshold, the utility is zero, because the considered ranges are too close to the \ac{MPAR} making the function useless. Then, the utility increases linearly as the range increases since it reaches its objective value, beyond which it saturates to 1. It is worth noticing that both the threshold and objective range are task depending parameters.

\subsection{Optimization algorithm}\label{sec:opt_alg}

To obtain a solution to the challenging and non-convex resource allocation problem defined in \eqref{eq_optimizationproblem} the iterative optimization algorithm in \cite{matlabref} is exploited. Therein, the interior-point approach to constrained optimization\footnote{Maximizing a utility is tantamount to minimizing the associated cost, given by the opposite of the utility.} is employed, which amounts to solve a sequence of approximate minimization problems which include non-negative constrained slack variables (as many as the inequality constrains of the original problem) and equality constraints. These are easier to solve than the original inequality-constrained problem and are handled either via a direct solution of the corresponding \ac{KKT} equations (via a linear approximation, i.e. a Newton step) or via a conjugate gradient method \cite{byrd1999interior, byrd2000trust, waltz2006interior}. Specifically, the algorithm first attempts to pursue a direct step. If it cannot be applied, it employs a conjugate gradient approach. Notably, one relevant case where the direct step is not exploited arises when the approximate problem is not locally convex near the current iterate. 

From an implementation point-of-view, the solution algorithm is based on the availability of an oracle (realized via a tailored numerical procedure) that provides the values for the objective function for each choice of the parameters as well as with the desired accuracy. This is indeed possible thanks to the analytic expressions which in implicit form rule the relationships among the objective and the different design parameters. 

It is fundamental to remark that no optimality claims can be done being the problem at hand \ac{NP} hard, in general. Nevertheless, the proposed technique leads to a solution that is a-posteriori practically effective, as shown by the results reported in Section \ref{sec_study_case}.

\section{Case studies}\label{sec_study_case}

In this section, some case studies for the pondered \ac{MPAR} system performing both search and \ac{COM} operations are analyzed. Specifically, the resource allocation is done after defining the priority weight for each task as well as the overall \ac{PAP} available at the system. Problem \eqref{eq_optimizationproblem} is solved using the Mathworks Matlab\textsuperscript{\textregistered} \emph{Quality-of-Service Optimization for Radar Resource Management} \cite{matlabopt} which performs a constrained minimization of a given objective function. The focus is on a scenario involving seven different tasks: three refer to search in \ac{LOS} scenarios (shortly referred to as Horizon, Long-range, and High-elevation, respectively), three \ac{COM}s with three different users, and a \ac{RIS}-aided search to tackle a \ac{NLOS} surveillance.

\subsection{Parameter setting}\label{sec_parameters}

Tests conducted in this paper refer to a \ac{MPAR} operating in X-band with its central frequency $f_0 = 10$ GHz. Now,
before providing the definition of all the involved parameters, for each considered task, the antenna coverage sector is specified in terms of angle limits, and observation range. In particular, the angular parameter setup specifies the following sector limits:
\begin{itemize}
\item Horizon, $[-45, 45]$ degrees in azimuth and $[0, 4]$ degrees in elevation,
\item Long-range, $[-30, 30]$ degrees in azimuth and $[0, 30]$ degrees in elevation, 
\item High-elevation, $[-45, 45]$ degrees in azimuth and $[30, 45]$ degrees in elevation.
\item \ac{COM} functions, $[-45, 45]$ degrees in azimuth and $[0, 45]$ degrees in elevation.
\item \ac{RIS}-aided, $[15, 20]$ degrees in azimuth and $[28, 32]$ degrees in elevation.
\end{itemize}

Additionally, the maximum range of interest (a.k.a. range limit) for each task is set as:
\begin{itemize}
\item Horizon, $40$ km,
\item Long-range, $70$ km, 
\item High-elevation, $50$ km.
\item \ac{COM} user 1, $45$ km,
\item \ac{COM} user 2, $55$ km,
\item \ac{COM} user 3, $65$ km,
\item \ac{RIS}-aided, $4$ km.
\end{itemize}

Other parameters for the three search tasks are summarized in Table \ref{tab_parameters_radar}, for the three \ac{COM} tasks are reported in Table \ref{tab_parameters_com}, and for the \ac{RIS}-aided (a uniform rectangular \ac{RIS} is considered during the analysis) search in Table \ref{tab_parameters_RIS}. It is worth highlighting that, a practical example for a search radar, which in part agrees with Table \ref{tab_parameters_radar}, is that of a ground surveillance system SHORAD (short range air defence) for air reconnaissance. In fact, it can possibly transmit with a low effective radiated power, and can also operate above C-band, where free-space loss is high \cite{arkoumaneas1982effectiveness}. Finally, in all the conducted simulations herein presented, $\emph{\text{PAP}}_{\text{min}_i}$, $i=1,\ldots, L$, is set to $0$ W$\cdot$m$^2$ unless otherwise stated.

\begin{table}[ht!]
\caption{\ac{LOS} search tasks simulation parameters.}
\centering
\begin{tabular}{cccc}
parameter & \multicolumn{3}{c}{value}\\
\hline
\hline
& Horizon & Long-range & High-elevation\\
\hline
$t_f$ (s) & $0.5$ & $6$ & $2$\\
$T_s$ (K) & $913$ & $913$ & $913$\\ 
$v_r$ (m/s) & $250$ & $250$ & $250$ \\
$\sigma$ (m$^2$) & $1$ & $1$ & $1$\\
$P_{fa}$ & $10^{-6}$ & $10^{-6}$ & $10^{-6}$\\
$L_s^{\text{LOS}}$ (dB) & $22$ & $19$ & $24$\\
$L_{\text{steer}}^{\text{LOS}}$ (dB) & $0.01$ & $0.13$ & $2.31$\\
\hline
\hline
\end{tabular}
\label{tab_parameters_radar}
\end{table}

\begin{table}[ht!]
\caption{\ac{COM} tasks simulation parameters.}
\centering
\begin{tabular}{cccc}
parameter & \multicolumn{3}{c}{value}\\
\hline
\hline
& user 1 & user 2 & user 3\\
\hline
$T_s^{\text{COM}}$ (K) & $916$ & $916$ & $916$\\ 
$B^{\text{COM}}$ (MHz) & $40$ & $40$ & $40$\\
$A_e^{\text{rx},k}$ (m$^2$) & $0.7\times 10^{-3}$ & $0.7 \times 10^{-3}$ & $0.7 \times 10^{-3}$\\
$L_s^{\text{COM}}$ (dB) & $27$ & $27$ & $27$\\
$L_{\text{steer}}^{\text{COM}}$ (dB) & $0.15$ & $0.62$ & $0.87$\\
\hline
\hline
\end{tabular}
\label{tab_parameters_com}
\end{table}

\begin{table}[ht!]
\caption{\ac{RIS}-aided search task simulation parameters.}
\centering
\begin{tabular}{cc}
parameter & value\\
\hline
\hline
$t_f$ (s) & $2$\\
$T_s$ (K) & $913$\\
$v_r$ (m/s) & $50$ \\
$\sigma$ (m$^2$) & $0.02$\\
$P_{fa}$ & $10^{-6}$\\
$L_s^{\text{NLOS}}$ (dB) & $19$\\
$G_{\text{patch}}$ (dB) & $4$\\ 
$\delta_x$, $\delta_y$ & $\lambda_0/2$\\
$N_1$, $N_2$ & $101$\\
$\eta_{\text{RIS}}$ & $0.8$\\
$r_{\text{NLOS}}$ (km) & $1$\\
$L_{\text{steer}}^{\text{NLOS}}$ (dB) & $1.25$\\
\hline
\hline
\end{tabular}
\label{tab_parameters_RIS}
\end{table}

\subsection{Case study 1}\label{sec_study1}

\begin{figure}[ht!]
\begin{center}
\subfigure[\ac{LOS} search]{\includegraphics[width=0.49\columnwidth]{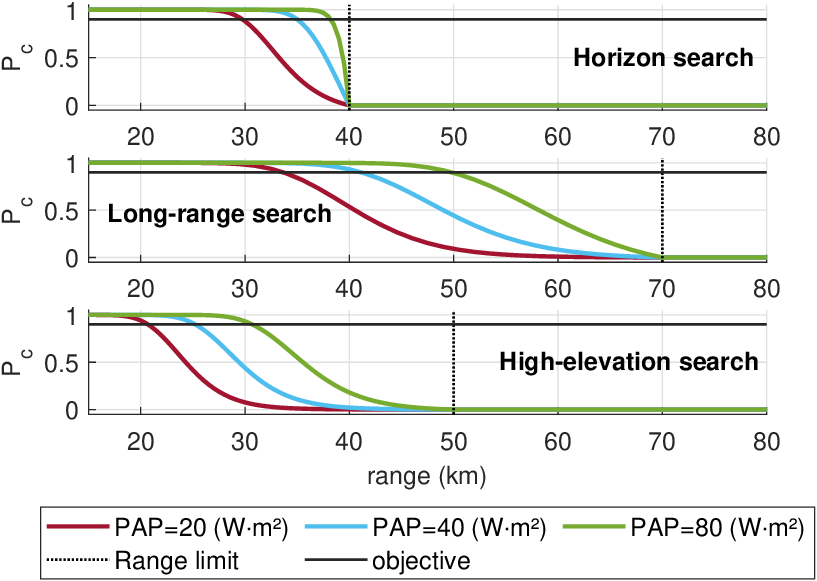}}
\subfigure[\ac{COM}]{\includegraphics[width=0.49\columnwidth]{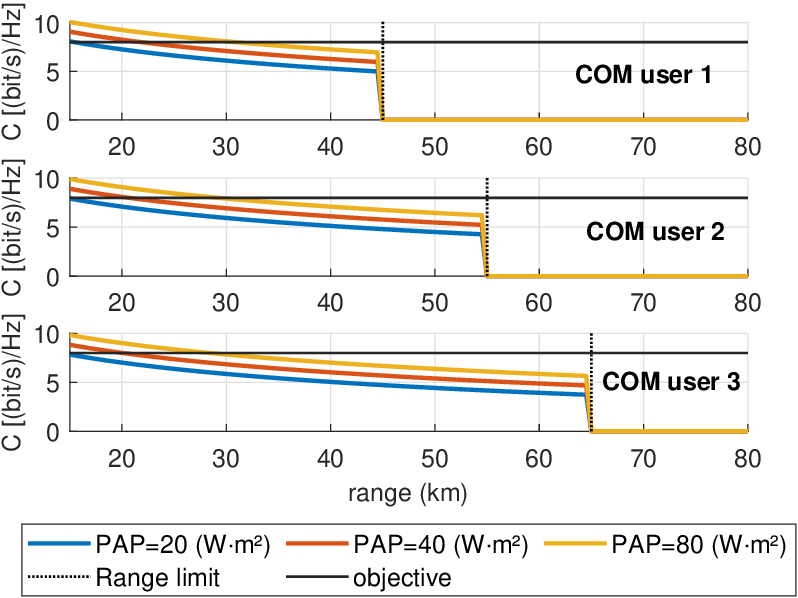}}
\subfigure[\ac{NLOS} search]{\includegraphics[width=0.49\columnwidth]{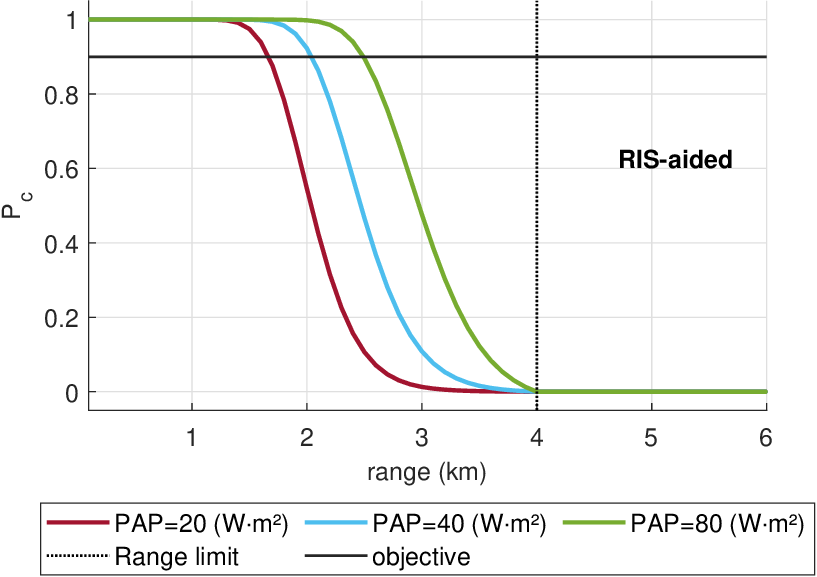}}
\end{center}
\caption{Cumulative \ac{$P_d$} for the \ac{LOS} (subfigure a) and \ac{NLOS} (subfigure c) search tasks, and channel capacity per bandwidth (subfigure b) for the \ac{COM} tasks, considering multiple per-task \ac{PAP} allocations.\label{fig1}}
\end{figure}

The first case study refers to a \ac{MPAR} with the parameters described in Section \ref{sec_parameters} assuming a SW1 fluctuating target model for both the high-speed targets considered in three \ac{LOS} search functions and for the small \ac{UAV} to be detected via  \ac{RIS}-aided surveillance. In this scenario, the cumulative \ac{$P_d$} \eqref{eq_cumulative} and channel capacity per bandwidth \eqref{eq_channelCap} are shown in Fig. \ref{fig1} versus range for three different values of the \ac{PAP} assigned to each task, viz. $[20, 40, 80]$ W$\cdot$m$^2$. Subfigures a) and c) of Fig. \ref{fig1} refer to search tasks, whereas subfigure b) to \ac{COM} operations.

\begin{figure}[ht!]
\begin{center}
\includegraphics[width=0.9\columnwidth]{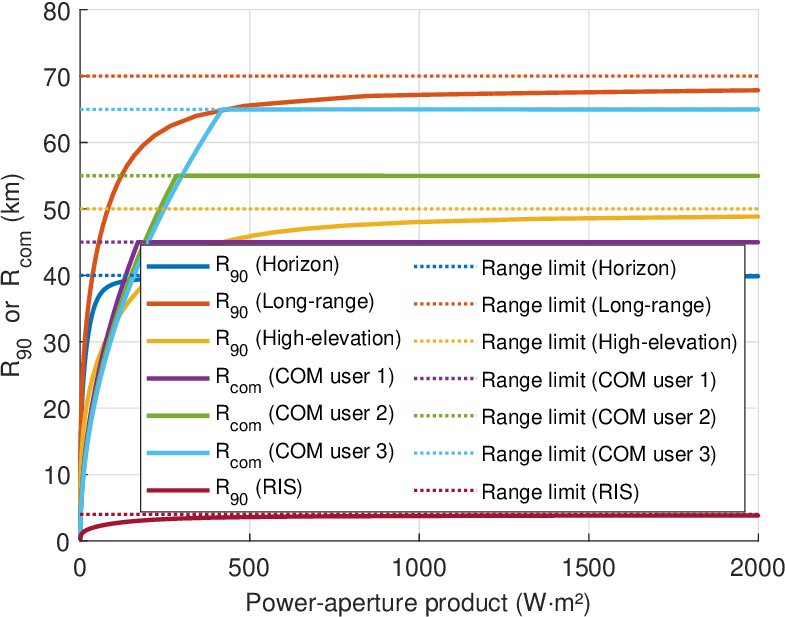}
\end{center}
\caption{Task quality versus assigned resource for the different radar operations.\label{fig2}}
\end{figure}

For all the subfigures of Fig. \ref{fig1}, the corresponding range limit is also shown. \ac{QoS} values beyond these limits are not of interest and set to zero as is evident for the \ac{COM} tasks. Moreover, the desired value for the cumulative \ac{$P_d$} (i.e., $P_{c_{\text{desired}}}=0.9$), and for the channel capacity per bandwidth (i.e., $C_{\text{desired}} = 8$ bit/s/Hz) are highlighted in the same graph. Hence, the corresponding range values $R_{90}$ and $R_{\text{com}}$ are derived for each \ac{PAP}s, numerically solving the equations $P_c^{\text{LOS}}(R_{\text{LOS}}|R_m)-P_{c_{\text{desired}}} = 0$, $P_c^{\text{NLOS}}(R_{\text{NLOS}}|R_m)-P_{c_{\text{desired}}} = 0$, and $C(R_{\text{COM}})-C_{\text{desired}}=0$ with respect to the variable $R_{\text{LOS}}$, $R_{\text{NLOS}}$, and $R_{\text{COM}}$, respectively. These results are reported in Fig. \ref{fig2}, where the task quality is shown versus the allocated resource to any specific task, i.e., $\emph{\text{PAP}}_i = \emph{\text{PAP}}_h$, for any  $i,h=1,\ldots,7$. As expected, increasing the assigned \ac{PAP} produces a growth of the task quality until its limit is attained. This means that if the current value of \ac{PAP} for a specific task is such that the range limit is almost attained, it is no longer required to allocate additional resource, since it does not produce appreciable improvements in the corresponding quality.

\begin{figure}[ht!]
\begin{center}
\subfigure[\ac{LOS} search]{\includegraphics[width=0.49\columnwidth]{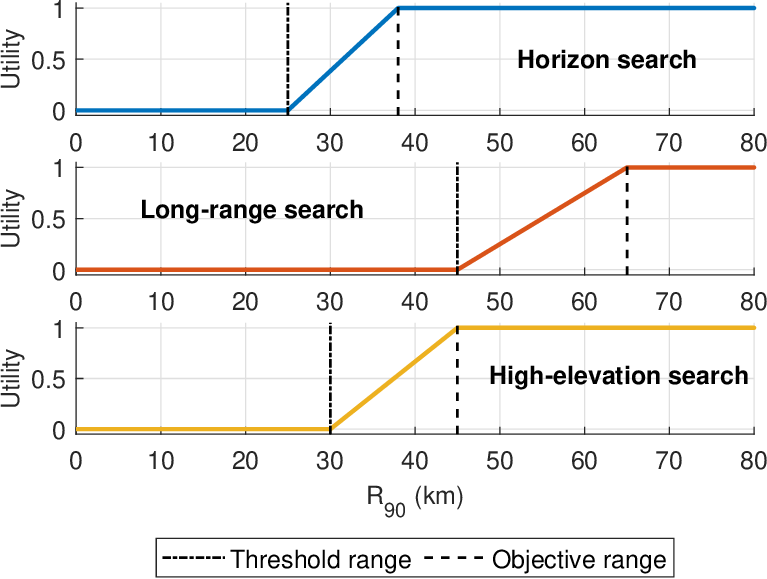}}
\subfigure[\ac{COM}]{\includegraphics[width=0.49\columnwidth]{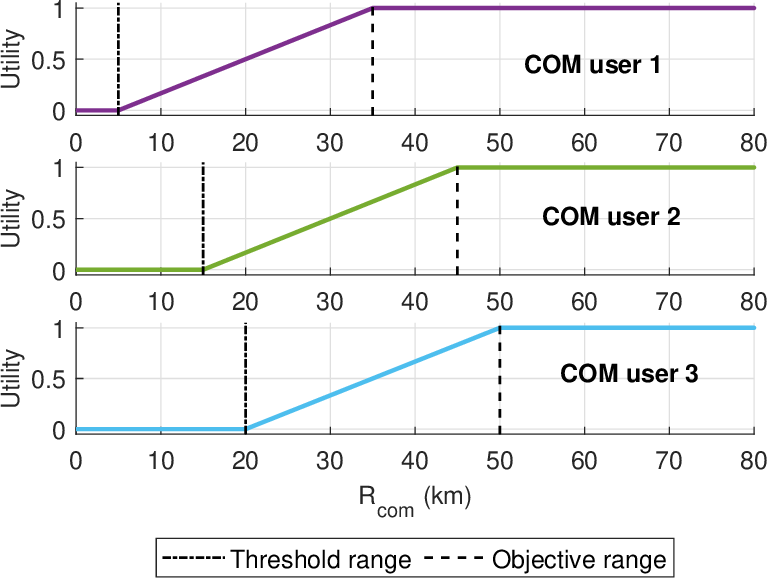}}
\subfigure[\ac{NLOS} search]{\includegraphics[width=0.49\columnwidth]{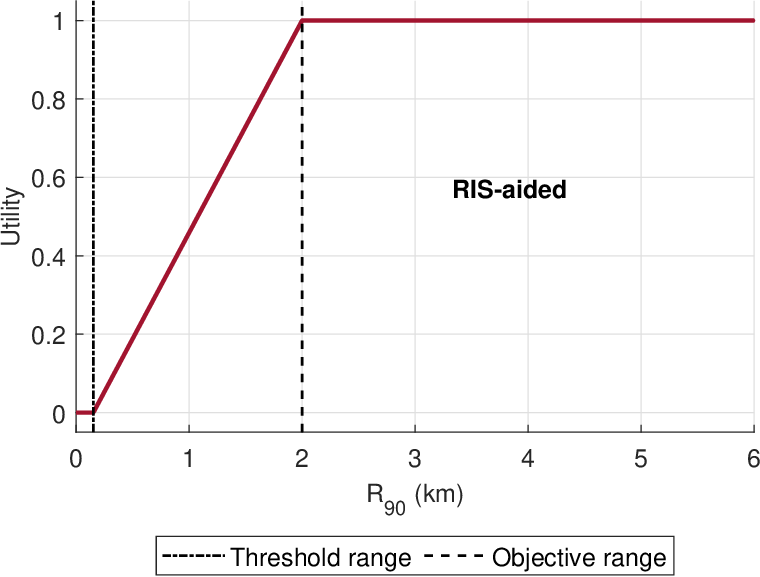}}
\end{center}
\caption{Utility functions for \ac{LOS} search tasks (subfigure a), \ac{COM} (subfigure b) and \ac{NLOS} search operation (subfigure c) tasks.\label{fig3}}
\end{figure}

In Fig. \ref{fig3} the utility functions for the above considered tasks are reported, particularizing the general form given by \eqref{eq_utility_func} setting the objective ranges to $R_o = [38, 65, 45, 35, 45, 50, 2]$ km and the threshold ranges to $R_t = [25, 45, 30, 5, 15, 20, 0.153]$ km for the three search (subfigure a), three \ac{COM} (subfigure b) and \ac{RIS}-aided (subfigure c) tasks, respectively. Note that, the threshold ranges are set following different requisites for each task under study. Precisely, for the \ac{LOS} search functions, it is the minimum range beyond which the mission is considered failed, because the target is too close to the radar for successfully activate subsequent actions. As to \ac{COM} tasks, the communication is assumed valid within a specific segment between two circles centered at the radar location, i.e., with the user located beyond a minimum distance from the radar until the possible maximum range of interest. For the \ac{RIS}-aided detection, the threshold range is set equal to the \ac{FFD} that can be computed as \cite{aubry2021reconfigurable}

\begin{equation}\label{eq_FFD}
R_{\text{FFD}} = \frac{2 \left(\max(\delta_x N_1, \delta_y N_2)\right)^2}{\lambda_0}.
\end{equation}

Therefore, for the parameter values summarized in Table \ref{tab_parameters_RIS}, \ac{FFD} computed via \eqref{eq_FFD} is approximately $153$ m. Finally, the objective ranges, that allow to reach the maximum utility, are set according to the mission requirements.

\begin{figure}[ht!]
\begin{center}
\includegraphics[width=0.9\columnwidth]{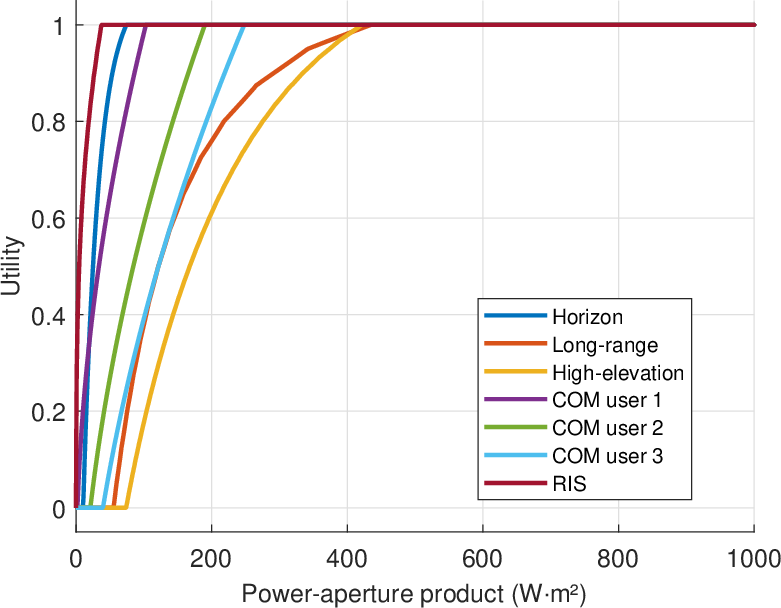}
\end{center}
\caption{Utility versus resource for the different radar operations.\label{fig4}}
\end{figure}

Moreover, using the above-described utility functions, the \ac{PAP} (namely, the resource) can be mapped to the utility space as shown in Fig. \ref{fig4}. From the inspection of these curves, it appears that the Long-range search, High-elevation search, \ac{COM} user 2 and 3 need to exploit non negligible \ac{PAP} values to reach non-zero utilities, viz., $56$, $74$, $22$, and $40$ W$\cdot$m$^2$, respectively. Conversely, the rest of the tasks are capable of reaching nonzero utilities with very low values of assigned \ac{PAP}. Moreover, the Long-range and High-elevation search functions demand high \ac{PAP} values to obtain the maximum utility, i.e., $422$ and $435$ W$\cdot$m$^2$. Interestingly, the operation that requires the minimum \ac{PAP} value to attain the maximum utility is the \ac{RIS}-aided search with \ac{PAP} of $38$ W$\cdot$m$^2$.

\begin{figure}[ht!]
\begin{center}
\subfigure[\ac{LOS} search]{\includegraphics[width=0.75\columnwidth]{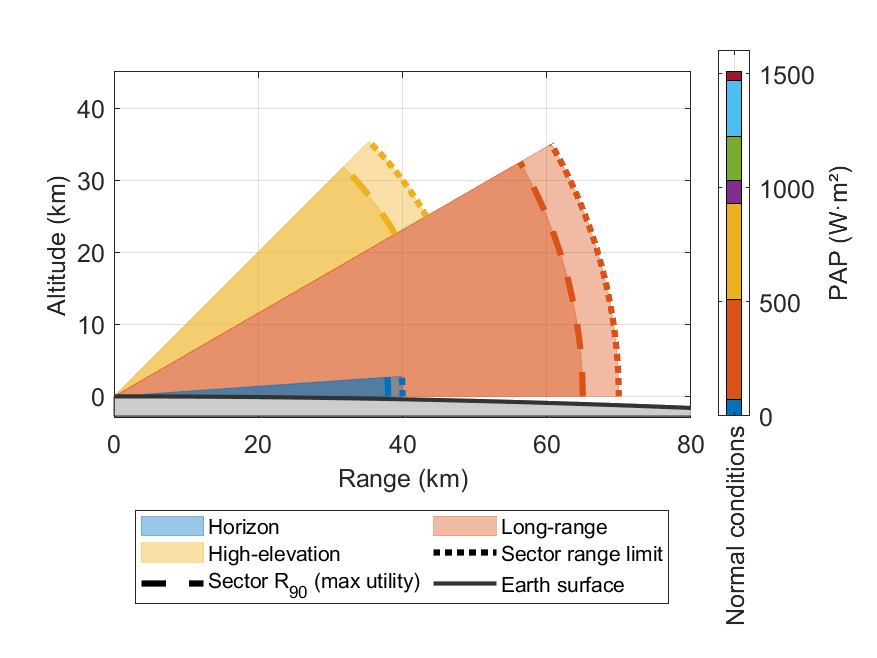}}
\subfigure[\ac{COM}]{\includegraphics[width=0.75\columnwidth]{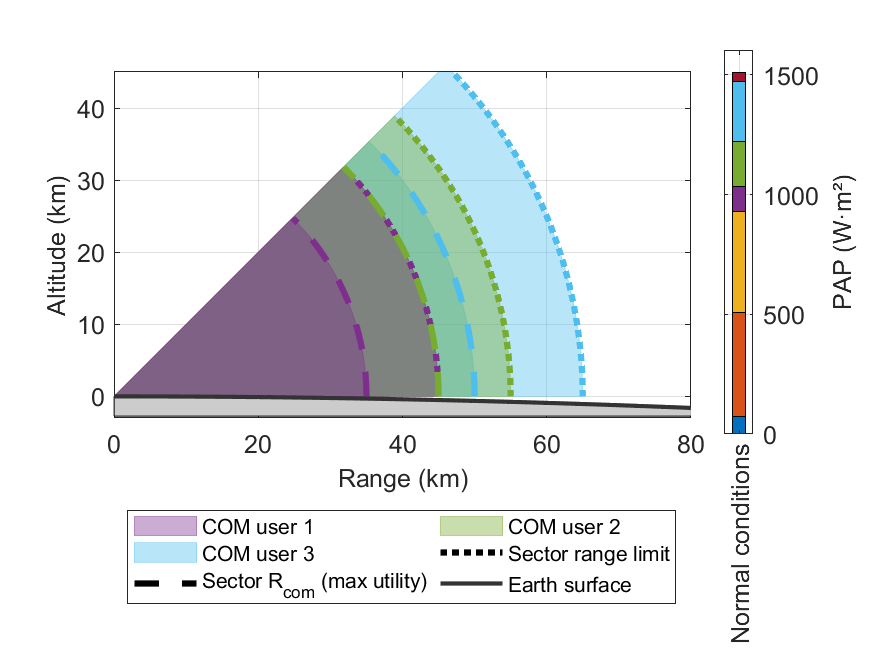}}
\subfigure[\ac{NLOS} search]{\includegraphics[width=0.75\columnwidth]{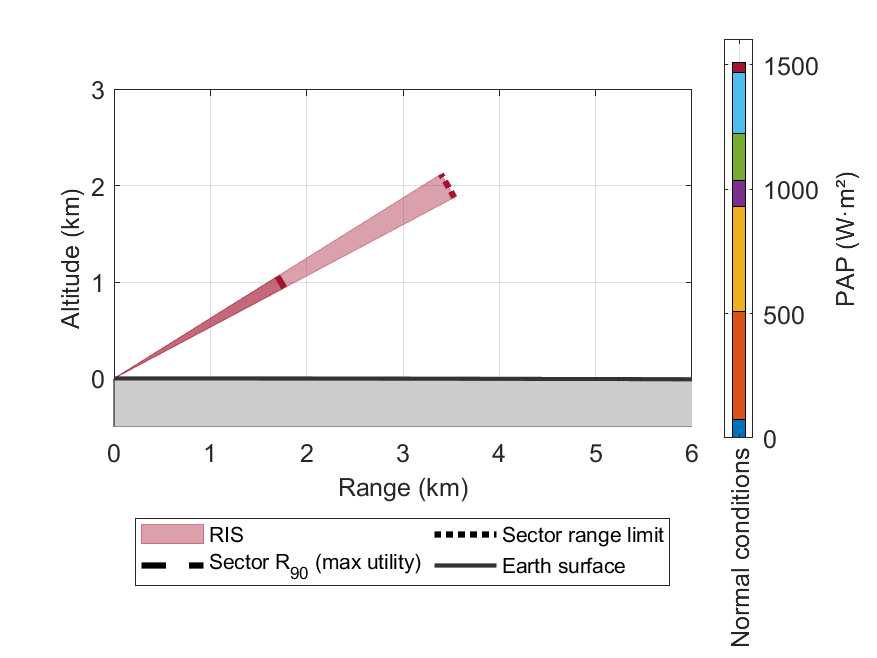}}
\end{center}
\caption{Resource allocation of \ac{MPAR} \ac{LOS} search tasks (subfigure a), \ac{COM} activities (subfigure b), and \ac{NLOS} search operation (subfigure c) under normal operational conditions. The line style of each attribute inherits the color of the corresponding task.\label{fig5}}
\end{figure}

Now, the first simulation analyzes the case where the resource allocation is performed under normal operational conditions (i.e., no optimization is performed) in which the maximum utility is reached for each of the operating tasks. Hence, each task exploits all the necessary resource (i.e., the maximum utility \ac{PAP}) to fulfill its demanded nominal objective, viz. cumulative \ac{$P_d$} and/or channel capacity per bandwidth. To highlight this distribution, Fig. \ref{fig5} proposes a graphical representation of the antenna coverage sectors as well as the objective value $R_{90}$ (respectively $R_{\text{com}}$) for the different radar operations. Subfigures refer to a) 
\ac{LOS} search, b) \ac{COM}, and c) \ac{NLOS} search tasks. Additionally, on the right side of this diagram a bar chart indicating the \ac{PAP} allocated to each task is also reported. Specifically, the maximum utility values are obtained with the allocation $\emph{\textbf{\text{PAP}}} = [74, 435, 422, 103, 190, 248, 37]^T$ W$\cdot$m$^2$, corresponding to a total \ac{PAP} used by the \ac{MPAR} of about $1509$ W$\cdot$m$^2$ (i.e., the sum of the maximum utility \ac{PAP} values for each task).

\begin{figure}[ht!]
\begin{center}
\subfigure[\ac{LOS} search]{\includegraphics[width=0.75\columnwidth]{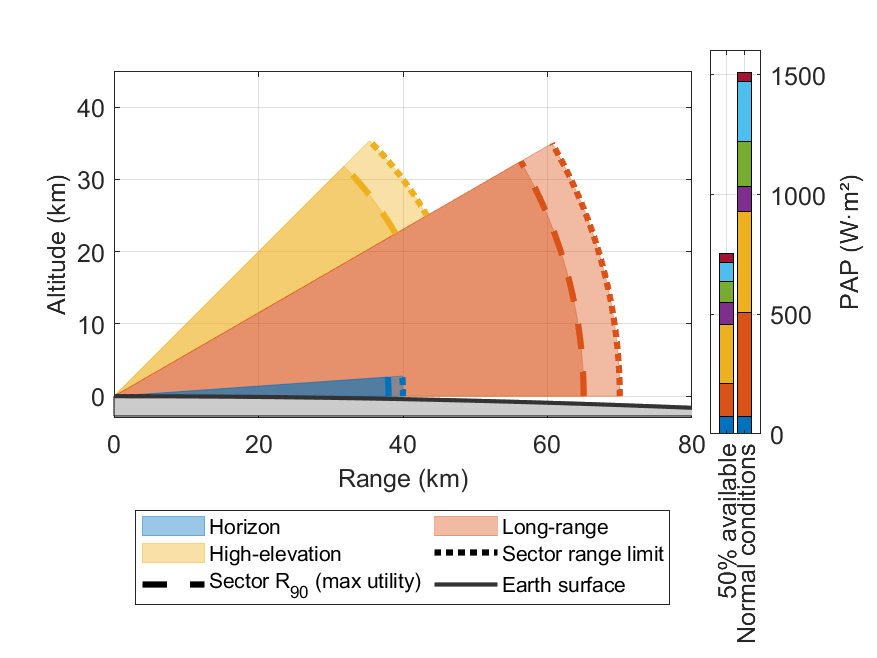}}
\subfigure[\ac{COM}]{\includegraphics[width=0.75\columnwidth]{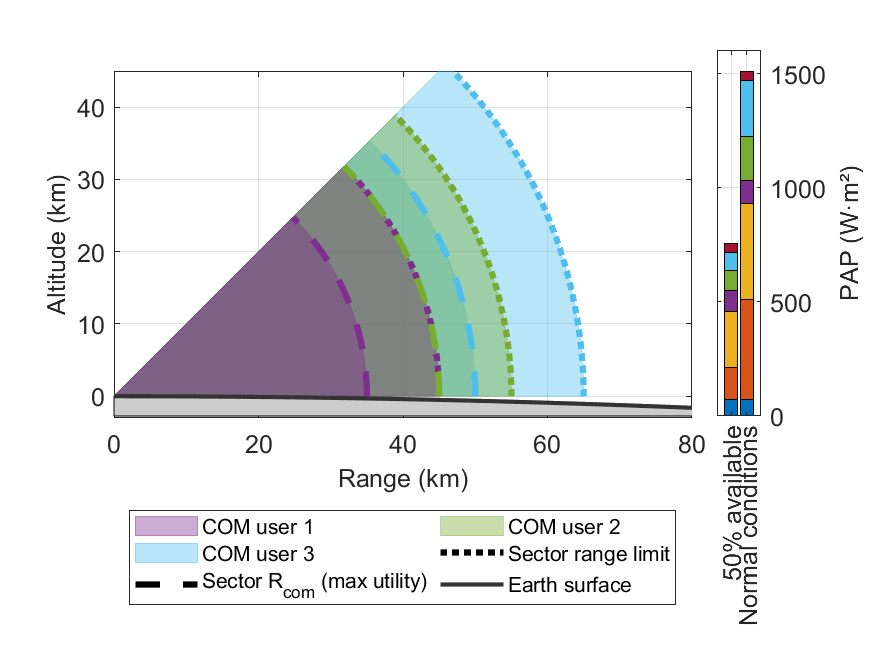}}
\subfigure[\ac{NLOS} search]{\includegraphics[width=0.75\columnwidth]{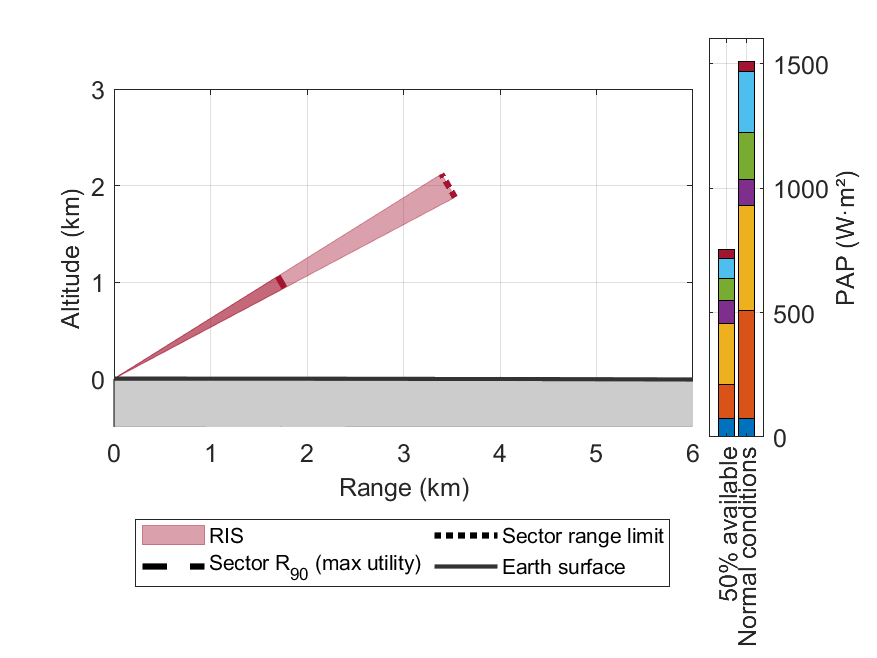}}
\end{center}
\caption{Resource allocation of \ac{MPAR} \ac{LOS} search tasks (subfigure a), \ac{COM} activities (subfigure b), and \ac{NLOS} search operation (subfigure c), with priority weights $\bw = [0.4, 0.1, 0.2, 0.06, 0.06, 0.06, 0.12]^T$.\label{fig6}}
\end{figure}

\begin{figure}[ht!]
\begin{center}
\subfigure[Horizon]{\includegraphics[width=0.45\columnwidth]{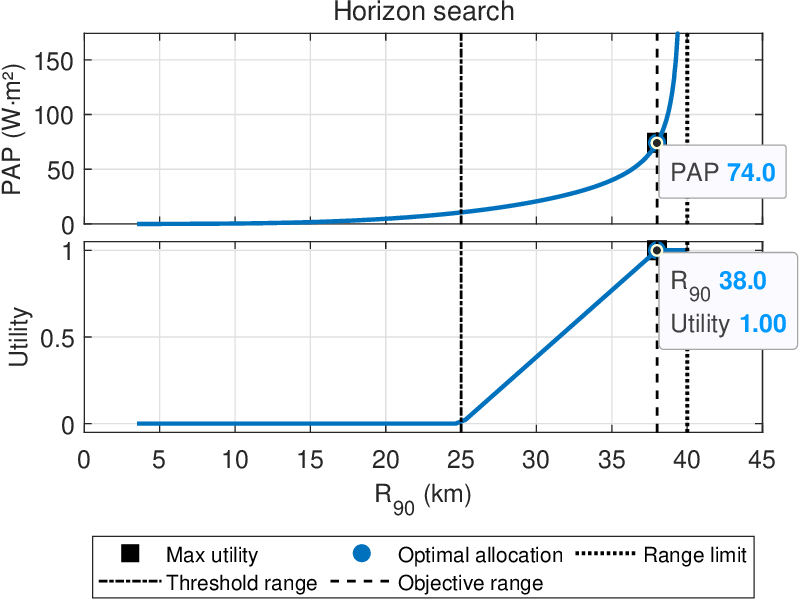}}
\subfigure[Long-range]{\includegraphics[width=0.45\columnwidth]{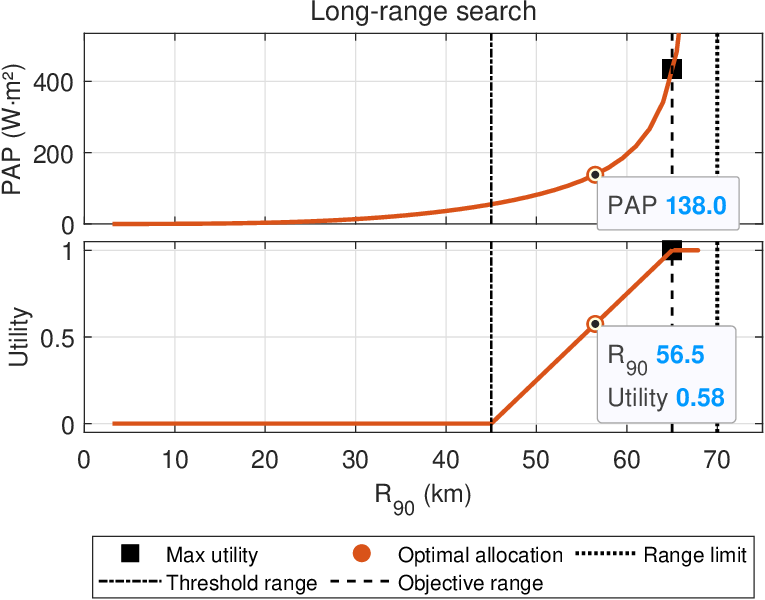}}\\
\subfigure[High-elevation]{\includegraphics[width=0.45\columnwidth]{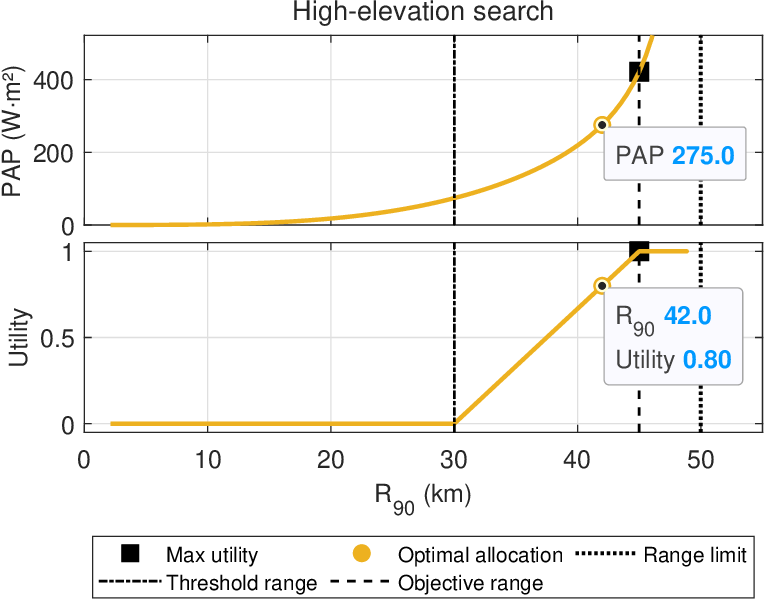}}
\subfigure[\ac{COM} user 1]{\includegraphics[width=0.45\columnwidth]{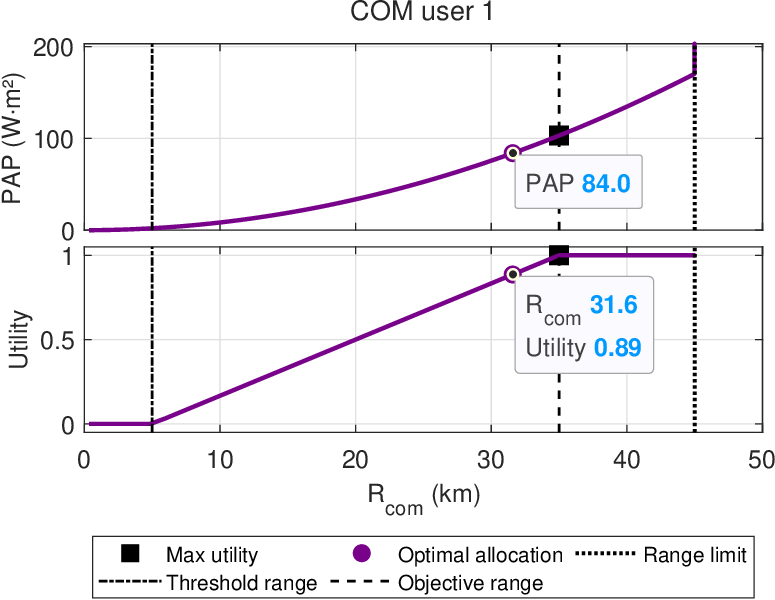}}\\
\subfigure[\ac{COM} user 2]{\includegraphics[width=0.45\columnwidth]{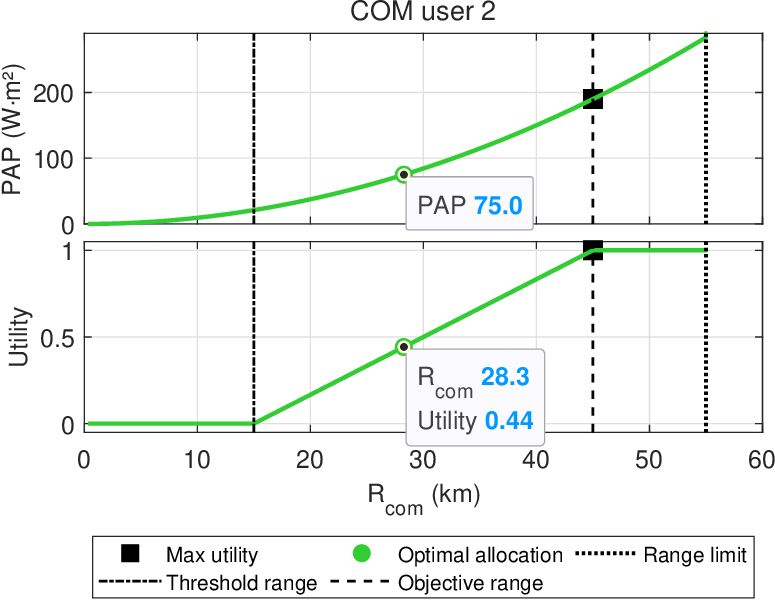}}
\subfigure[\ac{COM} user 3]{\includegraphics[width=0.45\columnwidth]{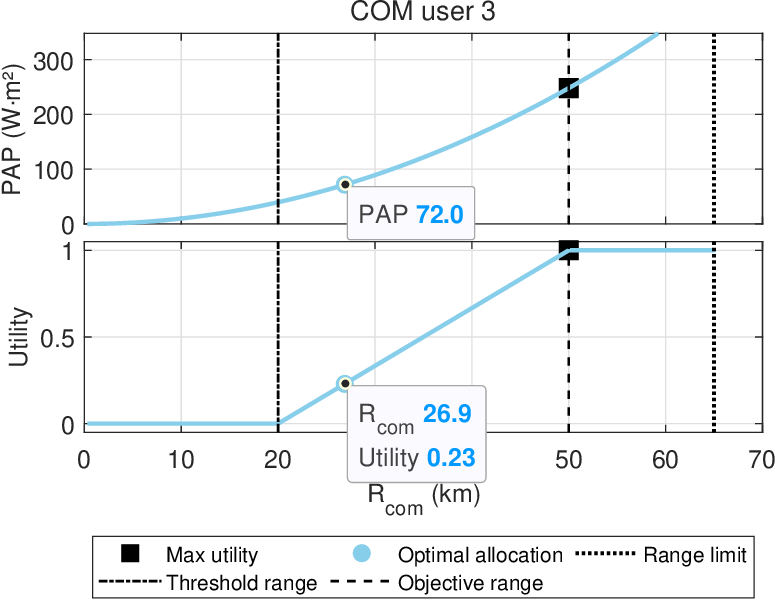}}\\
\subfigure[\ac{RIS}-aided]{\includegraphics[width=0.45\columnwidth]{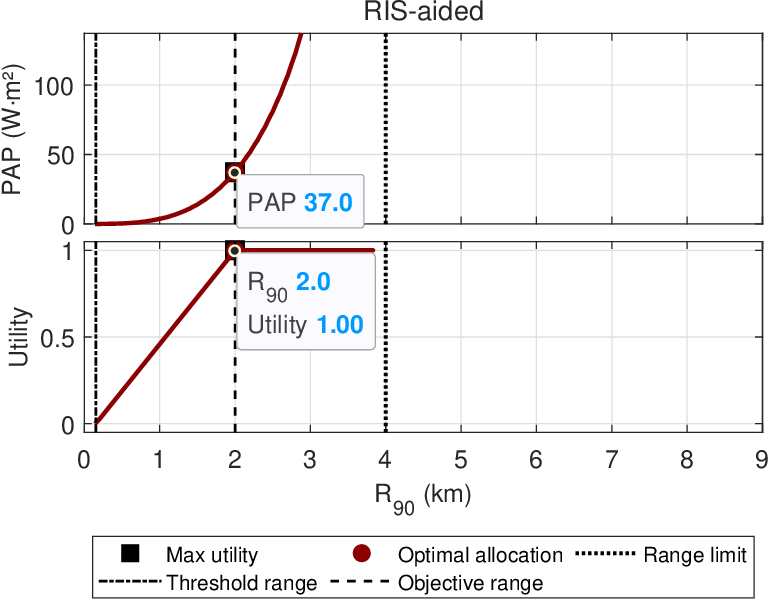}}
\end{center}
\caption{Optimized resource allocation and utility of \ac{MPAR} \ac{LOS} search (subfigures a-c), \ac{COM} (subfigures d-f), and \ac{NLOS} search (subfigure g) tasks, with priority weights $\bw = [0.4, 0.1, 0.2, 0.06, 0.06, 0.06, 0.12]^T$.\label{fig7}}
\end{figure}

Comparing the bar chart of Fig. \ref{fig5} with the diagram representing the utility versus resource of Fig. \ref{fig4}, it is evident that in the case of normal operational conditions, all tasks are capable of obtaining the maximum utility. In this situation, therefore, independently of the task, the respective quality metric is greater than or equal to its desired objective value. However, in some operating conditions, the total amount of resources available at the \ac{MPAR} cannot allow to assign the ideally required \ac{PAP} to each task. This can be also explained observing that, often, a non negligible part of the available resources should be reserved to other tasks (e.g., tracking)  \cite{barton2013radar}. For the above reasons, the \ac{RRM} should compute the optimal \ac{PAP} allocation, once its maximum available value is set. Hence, in this case study, the maximum \ac{PAP} is set to the $50\%$ of that under normal operational conditions, that is approximately $755$ W$\cdot$m$^2$. Moreover, the following set of priority weights is enforced, $\bw = [0.4, 0.1, 0.2, 0.06, 0.06, 0.06, 0.12]^T$, providing low priorities to \ac{COM} tasks with respect to search ones. Solving Problem \eqref{eq_optimizationproblem} with the above constraints results in the resource distribution reported in Fig. \ref{fig6}, where as before subfigures refer to a) \ac{LOS} search, b) \ac{COM}, and c) \ac{NLOS} search tasks. More specifically, the allocated \ac{PAP}s are equal to $\emph{\textbf{\text{PAP}}} = [74, 138, 275, 84, 75, 72, 37]^T$ W$\cdot$m$^2$. To give insights into the obtained results, Fig. \ref{fig7} shows for each task the optimal resource allocation in terms of \ac{PAP} versus the $R_{90}$ (respectively $R_{\text{com}}$) together with the corresponding utility, with subfigures referring to a)-c) \ac{LOS} search, d)-f) \ac{COM}, and g) \ac{NLOS} search operations. As expected, the \ac{RRM} allocates \ac{PAP} so that the maximum utility is reached for the Horizon search function, being the task with highest priority, with a corresponding $R_{90}=38$ km. Analogously, also the \ac{RIS}-aided search experiences an allocation of \ac{PAP} that allows to reach the maximum utility with $R_{90}=2$ km. This is because it has a medium priority (i.e., a weight $0.12$) together with the fact that it has low requirements in terms of resource. The worst case is observed in the \ac{COM} user 3 task where the \ac{PAP} allocation only ensures a utility of $0.23$, being its priority weight quite low and given by $0.06$.

\begin{figure}[ht!]
\begin{center}
\includegraphics[width=0.5\columnwidth]{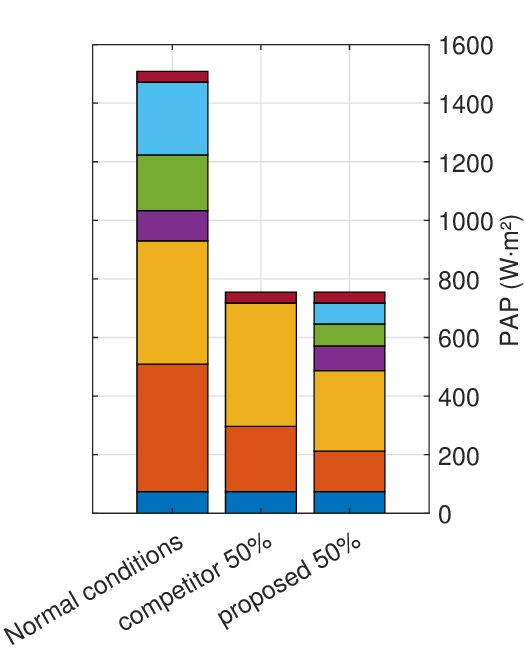}
\end{center}
\caption{Resource allocation of \ac{MPAR} \ac{LOS} search tasks, \ac{COM} activities, and \ac{NLOS} search operation, with priority weights $\bw = [0.4, 0.1, 0.2, 0.06, 0.06, 0.06, 0.12]^T$. Comparison of the proposed algorithm with the competitor maximizing the weighted sum of \ac{SNR}s.\label{fig02}}
\end{figure}

Now, the algorithm solving Problem \eqref{eq_optimizationproblem} with the weighted sum of \ac{SNR}s as objective function is considered as a possible competitor. In such a case, the optimization of the weighted \ac{SNR} together with the linear constraints gives rise to a linear programming problem. Results are graphically reported in Fig. \ref{fig02}, where the competitor provides a \ac{PAP} allocation, i.e., $\emph{\textbf{\text{PAP}}} = [74, 222, 421, 0, 0, 0, 37]^T$ W$\cdot$m$^2$, that substantially differs from that given by the proposed method, i.e., $\emph{\textbf{\text{PAP}}} = [74, 138, 275, 84, 75, 72, 37]^T$ W$\cdot$m$^2$. With the above allocation, the competitor reaches utilities equal to $1$, $0.81$, $1$, $0$, $0$, $0$, and $1$, for the seven tasks, respectively, with an average utility of $0.801$, whereas the proposed method provides as utility values 1, 0.58, 0.80, 0.89, 0.44, 0.23, and 1 with an average utility of $0.831$. Analyzing these results it is clear that the competitor does not allocate any resources to the COM tasks with a corresponding zero utility. Differently, the proposed method is capable of allocating some resources to all tasks providing at least some non-zero utilities. Moreover, the average utility reached by the proposed method is higher than that of the competitor (the competitor experiences a loss of $3.6\%$ in this case). Therefore, the validity and advantages of the proposed method should appear now much more evident.

To give further insights about the behavior of the proposed algorithm, the analysis of the case study 1 is repeated with a \ac{PAP} requirement set so as to ensure a utility of 0.5 for each task, viz. $\emph{\textbf{\text{PAP}}}_{\text{min}} = [25, 122, 168, 34, 85, 122, 5]^T$ W$\cdot$m$^2$. Solving Problem \eqref{eq_optimizationproblem} results in the resource distribution $\emph{\textbf{\text{PAP}}} = [74, 138, 245, 54, 85, 122, 37]^T$ W$\cdot$m$^2$, with corresponding utilities equal to 1, 0.58, 0.73, 0.68, 0.50, 0.50, and 1, for the seven tasks, respectively. In such a case, the average utility is $0.825$, whereas in the previous case without any guarantees on the minimum offered \ac{QoS} it was $0.831$. As expected enforcing additional requirements reduces the feasibility region (i.e., the available degrees of freedom) and possibly the resulting achieved objective function \eqref{eq_utility_total}. Moreover, from the inspection of these results, the evidence is that the \ac{RRM} allocates the \ac{PAP} so that the maximum utility is reached for the Horizon search function, being it the task with highest priority. Similarly, the \ac{RIS} is maintained invariant since it requires very low \ac{PAP}. However, the \ac{RRM}, accounting for a minimum ensured \ac{QoS} to the different tasks, tends to sacrifice the High-elevation task, and \ac{COM} user 1 that experience a loss in their achieved utility, to ensure that \ac{COM} user 2 and 3 attain the minimum required \ac{PAP} with utility 0.5. Definitely, when a non-zero lower bound on the \ac{PAP} is considered, the \ac{MPAR} is prone to subtract some resources to the (low weights) tasks whose allocation exceed the minimum requirements.

\begin{figure}[ht!]
\begin{center}
\includegraphics[width=0.85\columnwidth]{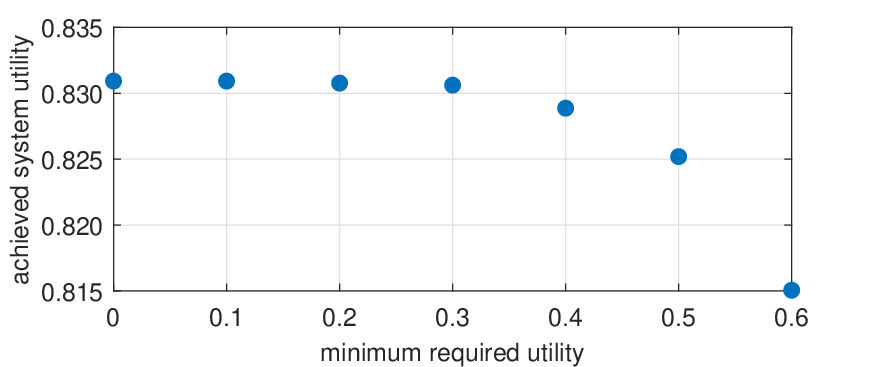}
\end{center}
\caption{Achieved objective function \eqref{eq_utility_total} versus minimum required utility (assumed equal for all the tasks).\label{fig15}}
\end{figure}

Before concluding this case study, Fig. \ref{fig15} shows the objective function \eqref{eq_utility_total} achieved by the proposed algorithm versus the utility value (assumed equal among the different tasks). As expected, the allocation performed by the \ac{RRM} attains a global utility that reduces as the constraints become more and more demanding.

\subsection{Case study 2}

\begin{figure}[ht!]
\begin{center}
\subfigure[\ac{LOS} search]{\includegraphics[width=0.75\columnwidth]{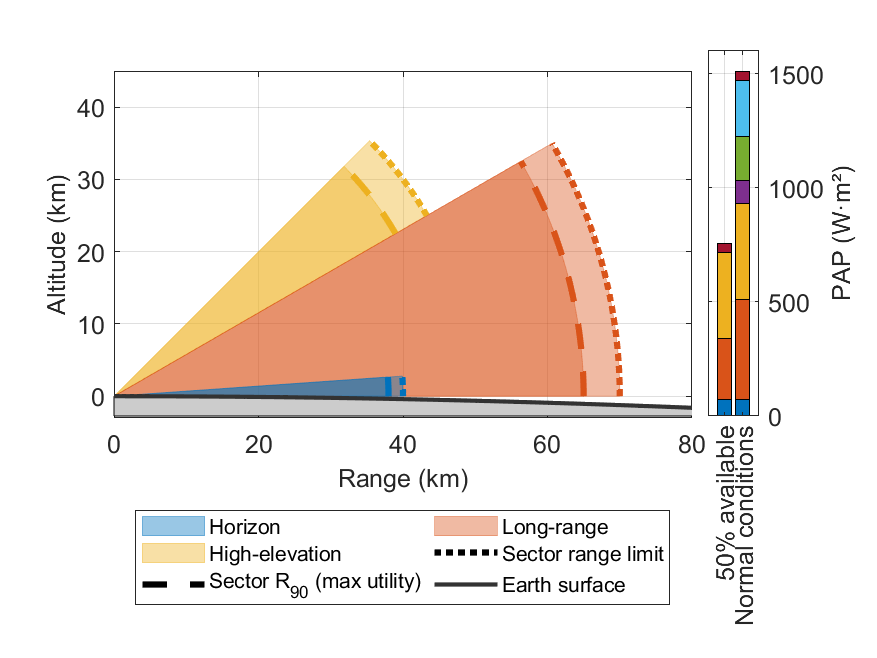}}
\subfigure[\ac{COM}]{\includegraphics[width=0.75\columnwidth]{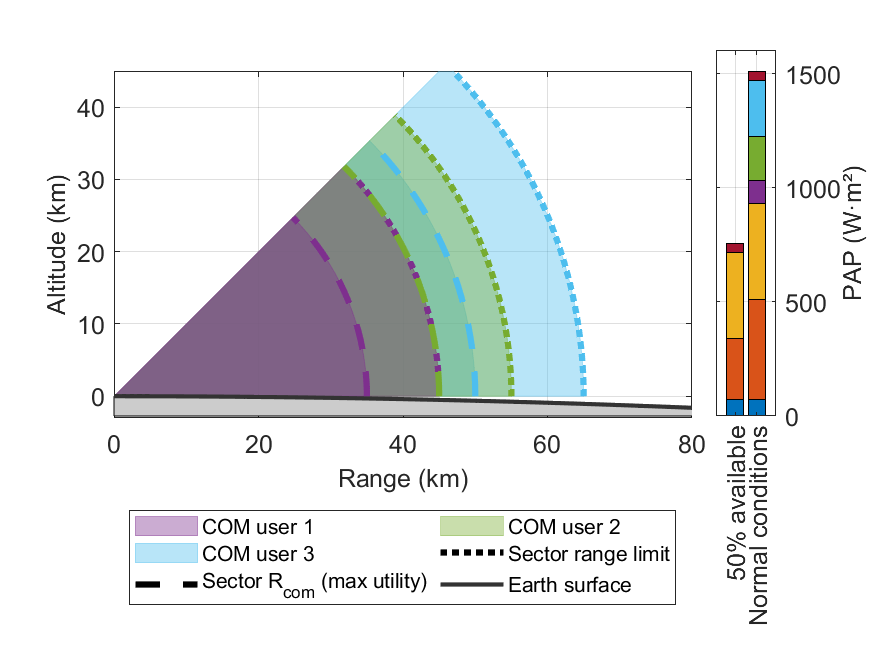}}
\subfigure[\ac{NLOS} search]{\includegraphics[width=0.75\columnwidth]{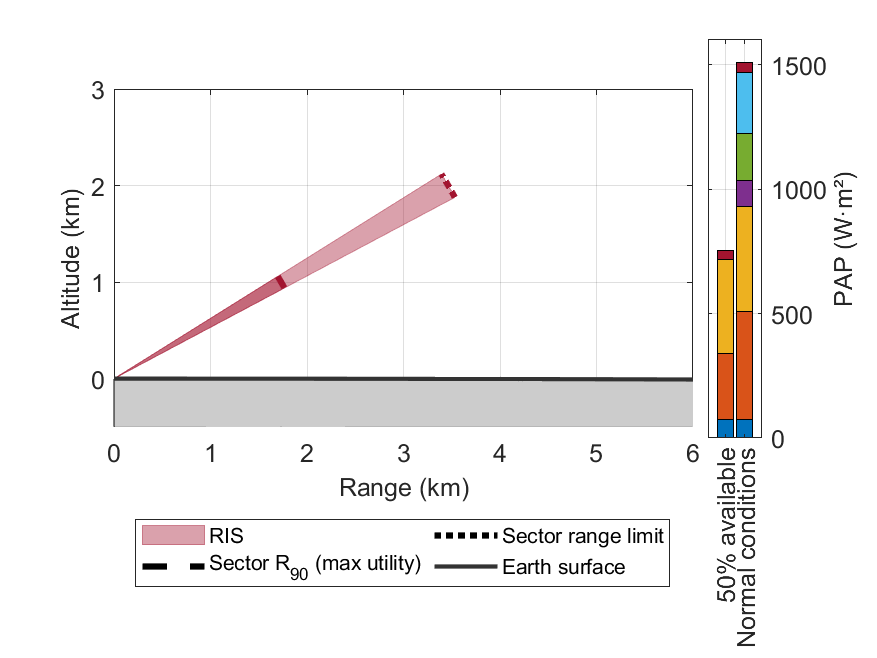}}
\end{center}
\caption{Resource allocation of \ac{MPAR} \ac{LOS} search tasks (subfigure a), \ac{COM} activities (subfigure b), and \ac{NLOS} search operation, with priority weights $\bw = [0.4, 0.20, 0.20, 0, 0, 0, 0.2]^T$.\label{fig6_pesinulli}}
\end{figure}

\begin{figure}[ht!]
\begin{center}
\subfigure[Horizon]{\includegraphics[width=0.45\columnwidth]{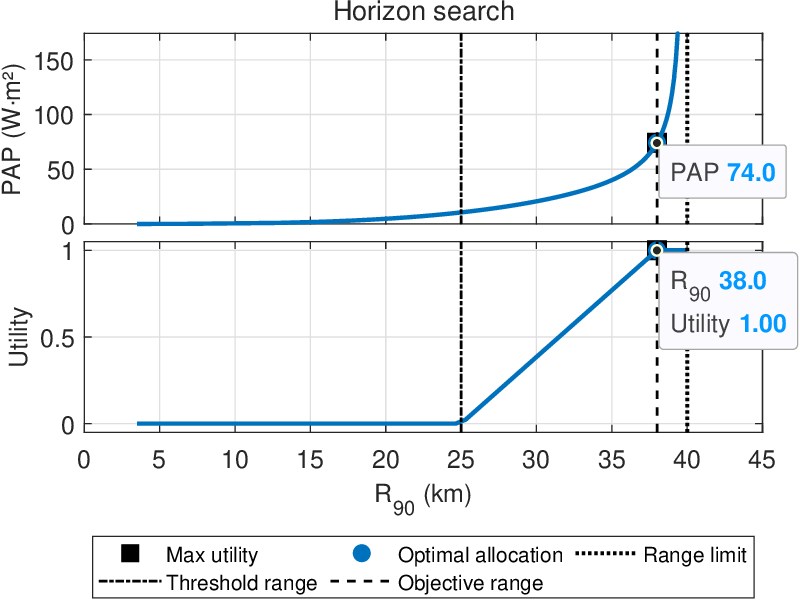}}
\subfigure[Long-range]{\includegraphics[width=0.45\columnwidth]{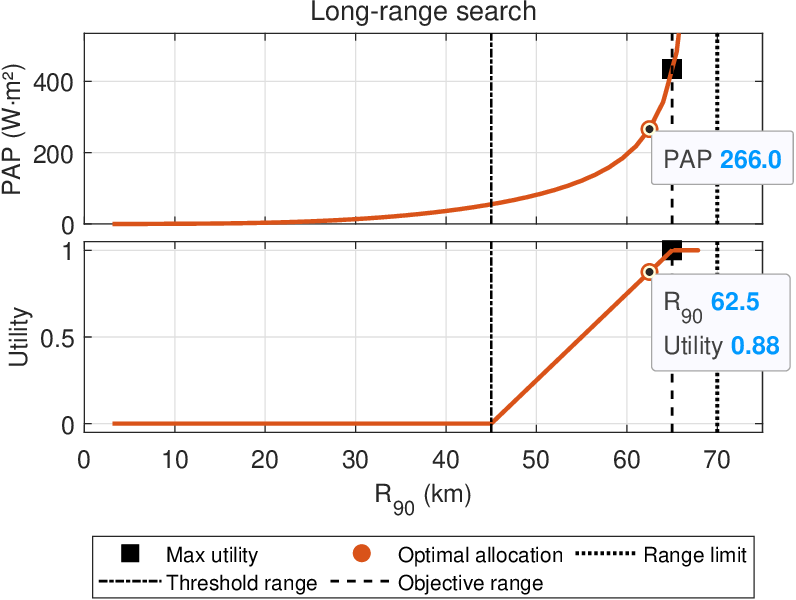}}\\
\subfigure[High-elevation]{\includegraphics[width=0.45\columnwidth]{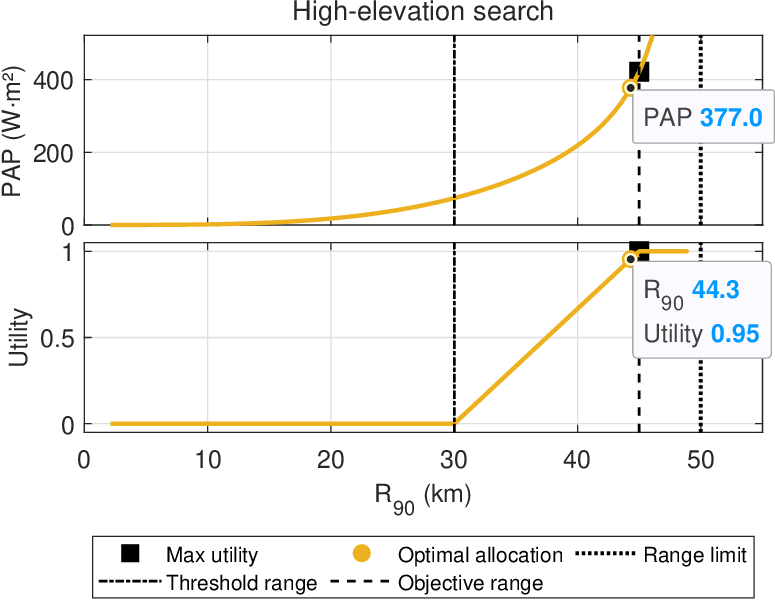}}
\subfigure[\ac{COM} user 1]{\includegraphics[width=0.45\columnwidth]{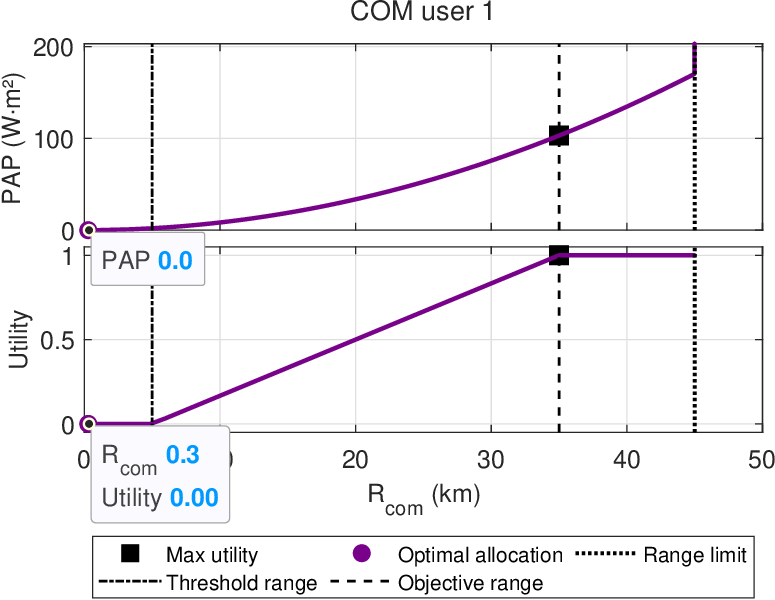}}\\
\subfigure[\ac{COM} user 2]{\includegraphics[width=0.45\columnwidth]{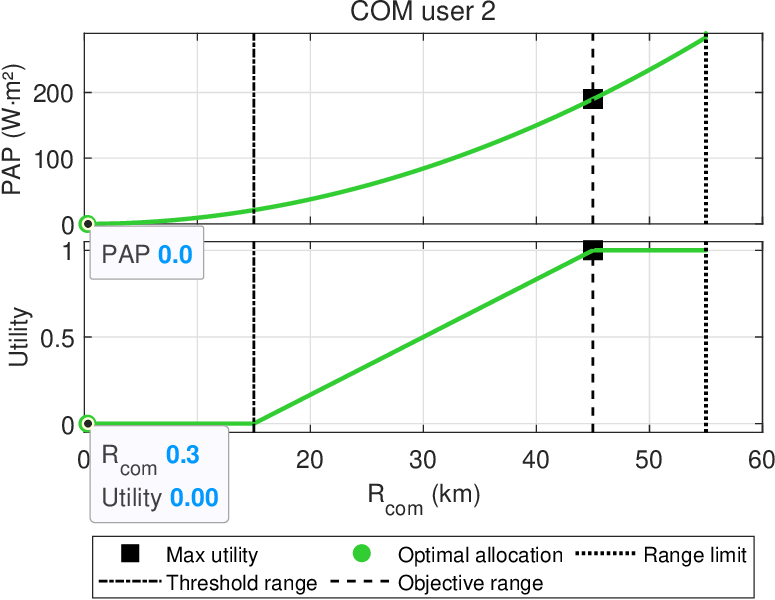}}
\subfigure[\ac{COM} user 3]{\includegraphics[width=0.45\columnwidth]{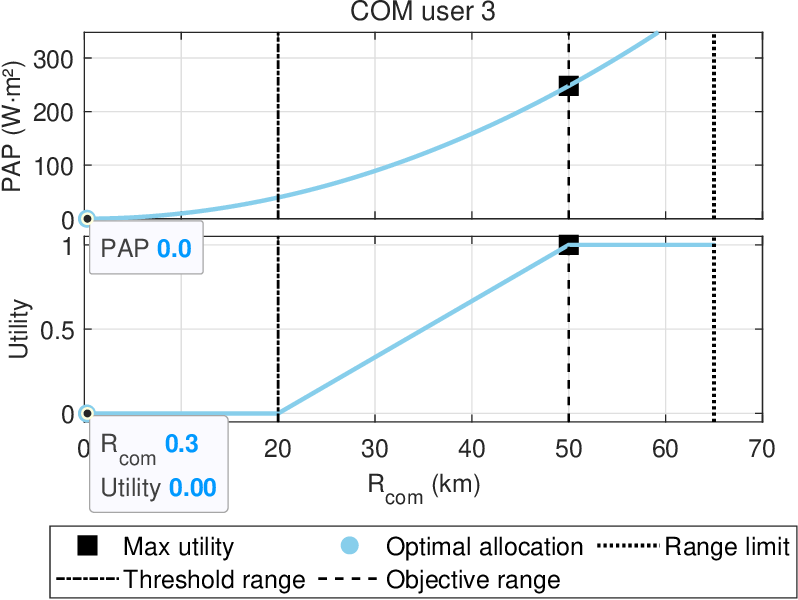}}\\
\subfigure[\ac{RIS}-aided]{\includegraphics[width=0.45\columnwidth]{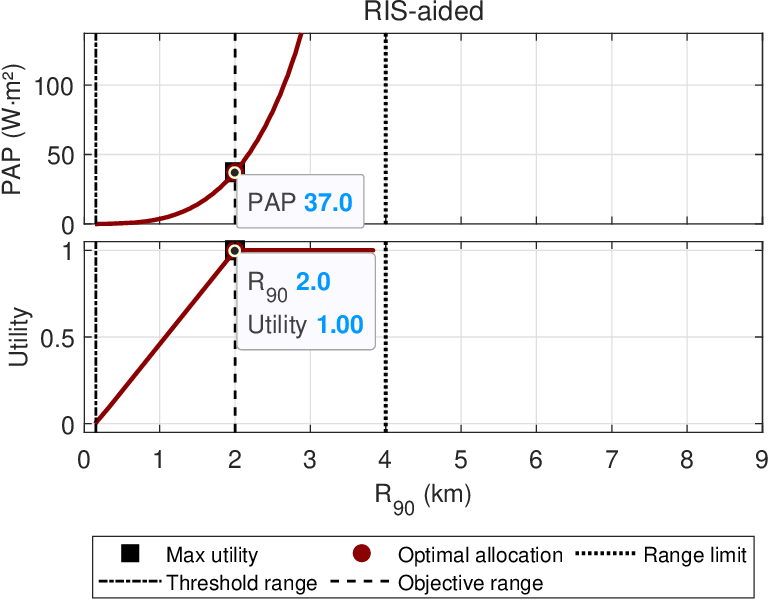}}
\end{center}
\caption{Optimized resource allocation and utility of \ac{MPAR} \ac{LOS} search (subfigures a-c), \ac{COM} (subfigures d-f), and \ac{NLOS} search (subfigure g) tasks, with priority weights $\bw = [0.4, 0.20, 0.20, 0, 0, 0, 0.2]^T$.\label{fig7_pesinulli}}
\end{figure}

In this situation, the \ac{PAP} allocation is performed for a different set of priority weights, again setting its maximum value to $755$ W$\cdot$m$^2$, i.e., half of that used under normal operational conditions. As a matter of fact, the priority weights for the \ac{COM} tasks are fixed to 0, resulting in the vector $\bw = [0.4, 0.2, 0.2, 0, 0, 0, 0.2]^T$. The solution to Problem \eqref{eq_optimizationproblem} with the above constraints produces the \ac{PAP} assignment over the considered tasks illustrated in Fig. \ref{fig6_pesinulli}, where subfigures refer to a) \ac{LOS} search tasks, c) \ac{COM} tasks, and d) \ac{RIS}-aided search task. Specifically, the allocated \ac{PAP} values are $\emph{\textbf{\text{PAP}}} = [74, 266, 378, 0, 0, 0, 37]^T$ W$\cdot$m$^2$. Again, Fig. \ref{fig7_pesinulli} shows for each task the optimal resource distribution in terms of \ac{PAP} versus $R_{90}$ (respectively $R_{\text{com}}$) together with the corresponding utility, with subfigures referring to a)-c) \ac{LOS} search tasks, d)-f) \ac{COM} tasks, and g) \ac{RIS}-aided search task. As expected the \ac{RRM} does not allocate any \ac{PAP} to the \ac{COM} tasks reflecting the associated zero priority weights. On the contrary, the Long-range and High-elevation experience a growth in the assignment of their resources, with a consequent increment of utility that increases from $0.65$ to $0.88$ and from $0.83$ to $0.95$ w.r.t. the case study 1, respectively. Obviously, the other two tasks (namely, Horizon and \ac{RIS}-aided search), having already reached their maximum utility, continue to maintain the same allocation as before.

\subsection{Case study 3}

The test performed in this subsection is devoted to the impact of the antenna pointing direction on the performance of the \ac{MPAR} in terms of resource distribution over the different tasks. In particular, for all tasks, the term accounting for scanning losses is fixed according to the values summarized in Table \ref{tab_parameters_worst}. Moreover, as to the other parameters, this study refers to the same simulation setting as in Section \ref{sec_study1}, apart for, as already specified losses accounting for the spatial selectivity of the antenna gain are set equal to their respective worst case for each angular sector.

\begin{table}[ht!]
\caption{Scanning loss (expressed in dB) for the worst antenna pointing direction case.}
\centering
\begin{tabular}{ccccc}
HHorizon & Long-range & High-elevation & \ac{COM} user 1-3 & \ac{RIS} \\
\hline
\hline
$0.02$ & $1.25$ & $3.01$ & $1.51$ & $7.45$\\
\hline
\hline
\end{tabular}
\label{tab_parameters_worst}
\end{table}

\begin{figure}[ht!]
\begin{center}
\subfigure[\ac{LOS} search]{\includegraphics[width=0.75\columnwidth]{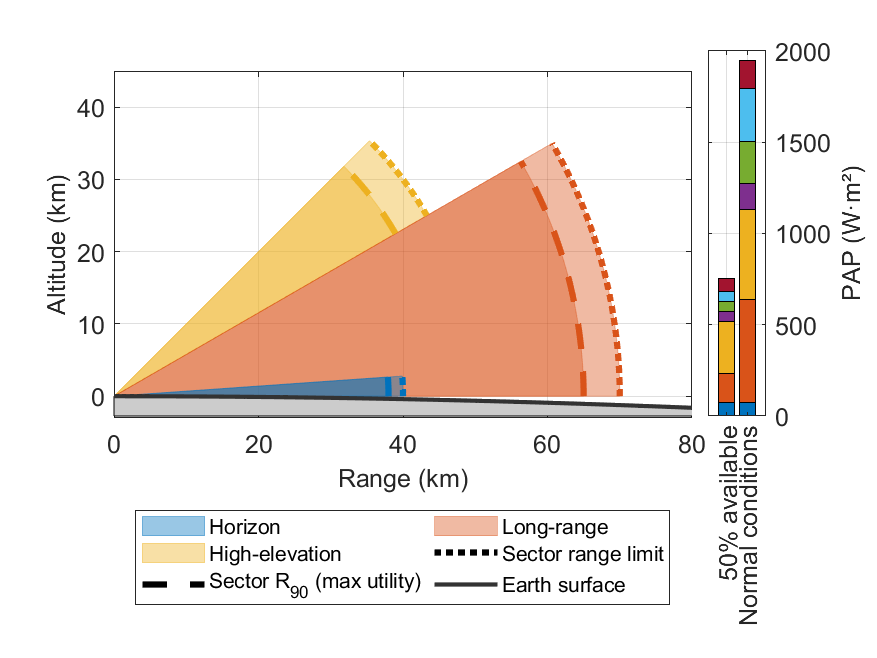}}
\subfigure[\ac{COM}]{\includegraphics[width=0.75\columnwidth]{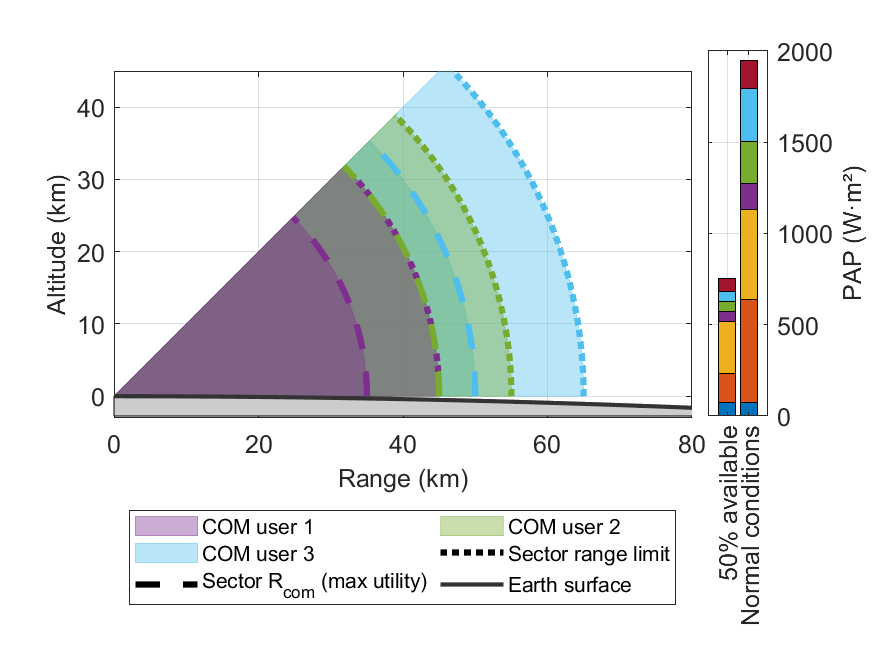}}
\subfigure[\ac{NLOS} search]{\includegraphics[width=0.75\columnwidth]{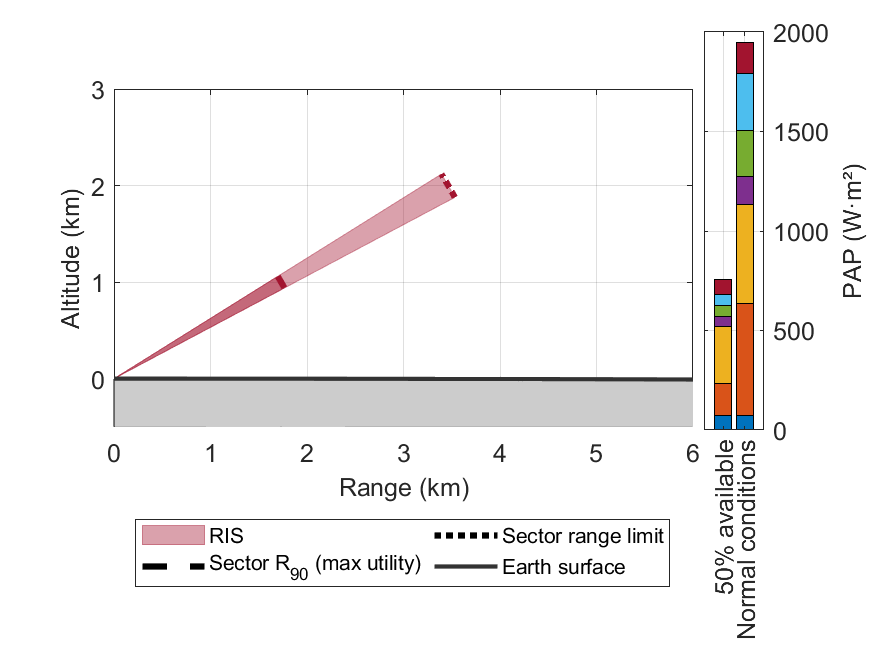}}
\end{center}
\caption{Resource allocation of \ac{MPAR} \ac{LOS} search tasks (subfigure a), \ac{COM} activities (subfigure b), and \ac{NLOS} search operation, assuming the worst case scanning loss.\label{fig6_worst}}
\end{figure}

\begin{figure}[ht!]
\begin{center}
\subfigure[Horizon]{\includegraphics[width=0.45\columnwidth]{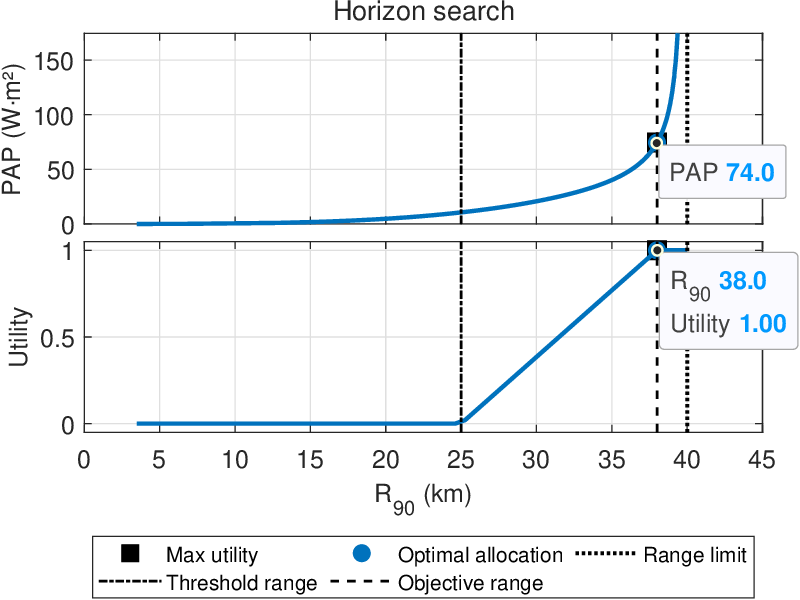}}
\subfigure[Long-range]{\includegraphics[width=0.45\columnwidth]{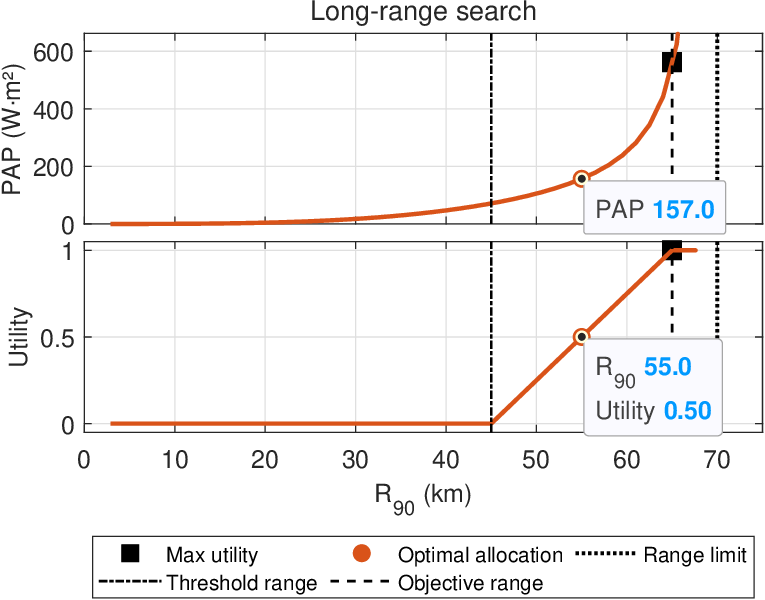}}\\
\subfigure[High-elevation]{\includegraphics[width=0.45\columnwidth]{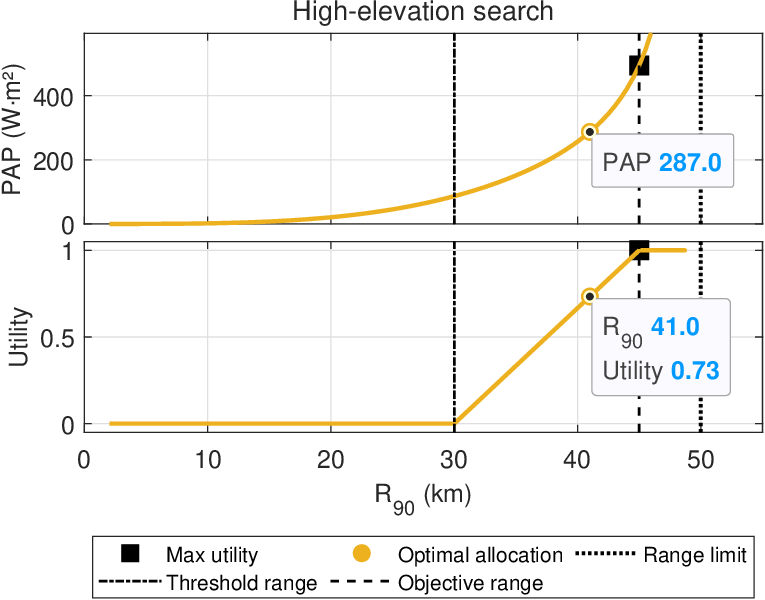}}
\subfigure[\ac{COM} user 1]{\includegraphics[width=0.45\columnwidth]{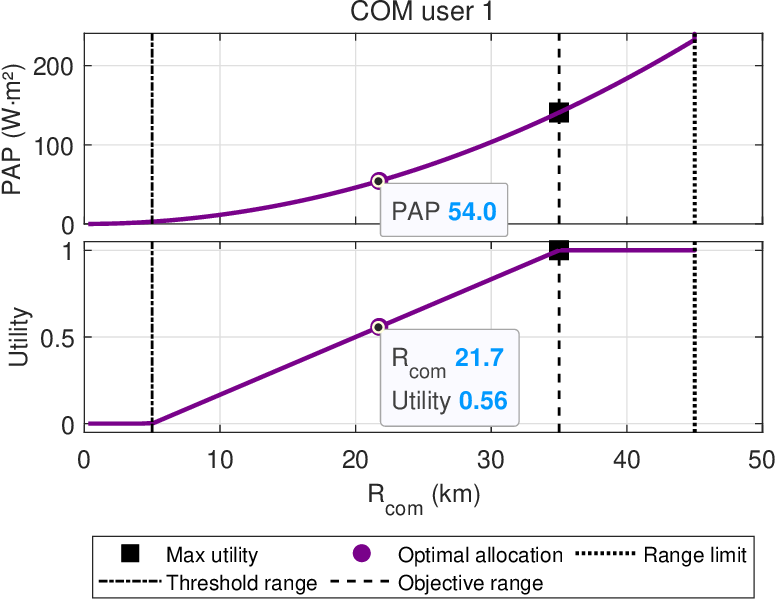}}\\
\subfigure[\ac{COM} user 2]{\includegraphics[width=0.45\columnwidth]{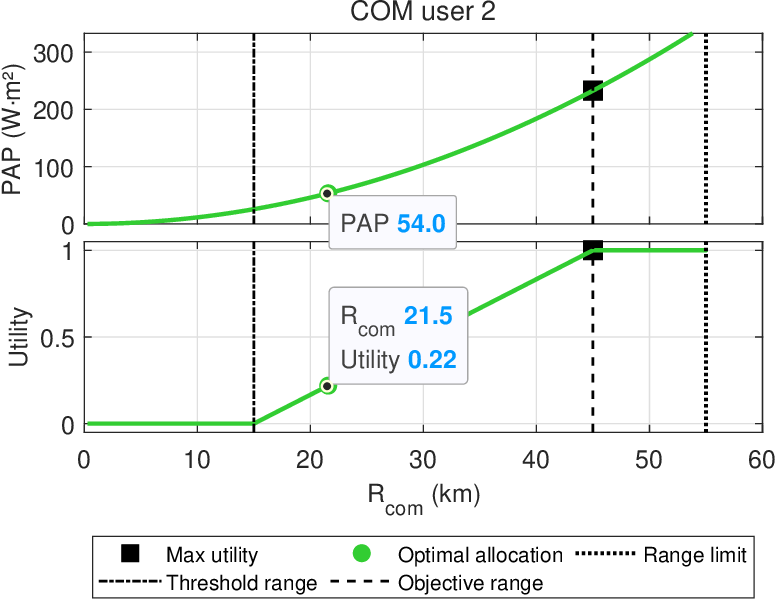}}
\subfigure[\ac{COM} user 3]{\includegraphics[width=0.45\columnwidth]{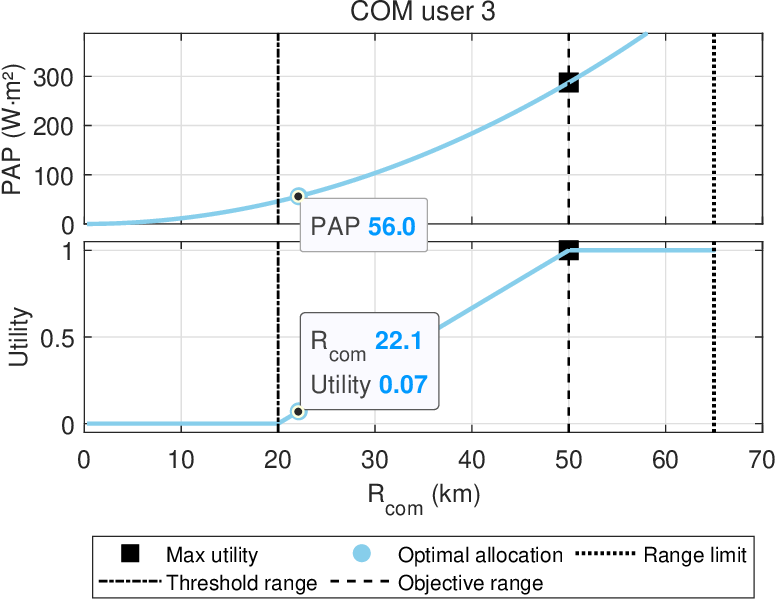}}\\
\subfigure[\ac{RIS}-aided]{\includegraphics[width=0.45\columnwidth]{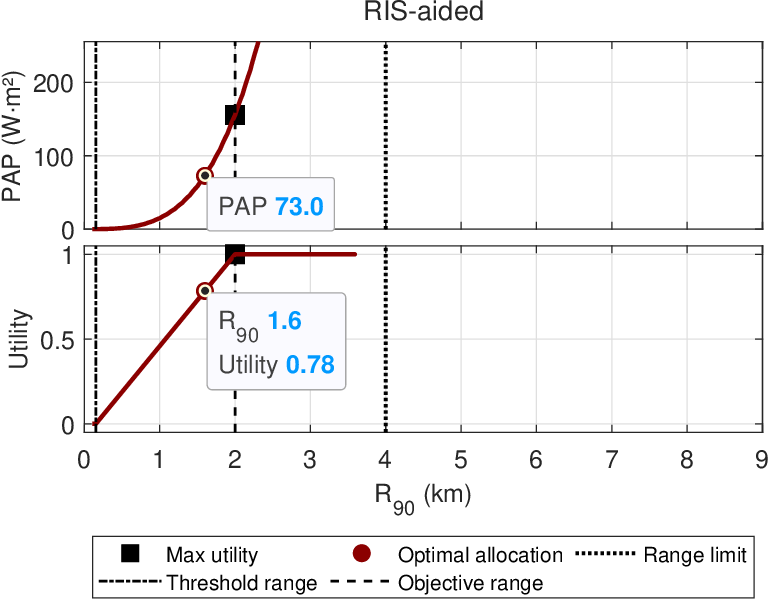}}
\end{center}
\caption{Optimized resource allocation and utility of \ac{MPAR} \ac{LOS} search (subfigures a-c), \ac{COM} (subfigures d-f), and \ac{NLOS} search (subfigure g) tasks, assuming the worst case scanning loss.\label{fig7_worst}}
\end{figure}

The conducted test considers the availability of maximum \ac{PAP} of $755$ W$\cdot$m$^2$ (that is again approximately the $50\%$ of that under normal operational conditions in the case study 1), with the same priority weights as in the first case study. Solving Problem \eqref{eq_optimizationproblem} with the above constraints results in the \ac{PAP} assignment illustrated in Fig. \ref{fig6_worst}, where subfigures refer to a) \ac{LOS} search, c) \ac{COM}, and d) \ac{RIS}-aided search tasks. More in detail, the allocated \ac{PAP}s are now equal to $\emph{\textbf{\text{PAP}}} = [74, 157, 287, 54, 54, 56, 73]^T$ W$\cdot$m$^2$, respectively. Again, to further shed light on the results, Fig. \ref{fig7_worst} shows for each task the optimal resource allocation in terms of \ac{PAP} versus $R_{90}$ (respectively $R_{\text{com}}$) along with their corresponding utility, with subfigures referring to a)-c) \ac{LOS} search, d)-f) \ac{COM}, and g) \ac{RIS}-aided search tasks. It is now interesting to observe that the resource allocation does not follow the trend as in the scenario analyzed in Section \ref{sec_study1}. In fact, the \ac{COM} tasks are all penalized with a reduction in the assignment of their \ac{PAP} due to their very low priorities (i.e., $0.06$). The majority of resources are allocated to the other tasks, with the Horizon search function that attains its maximum utility thanks to the attributed high priority. The \ac{RIS}-aided search task also reaches a high utility of 0.78 because of a joint combination of a medium priority weight and a reduced \ac{PAP} necessary to satisfy it. Finally, it is worth observing that all the considered tasks (except the Horizon) suffer the effect of the scanning loss that in turn reflects on a higher \ac{PAP} that is required to reach the same utility. Therefore, the \ac{RRM} tends to sacrifice the tasks with the lowest priority, i.e., \ac{COM} ones, to guarantee sufficient performance to the others.

\section{Concluding remarks}\label{sec_concl}

This paper has addressed the problem of optimal \ac{PAP} allocation in a \ac{MPAR} system performing \ac{ISAC} operations. More specifically, the considered methodology has been aimed at solving the \ac{QoS} optimization problem jointly accounting for search scenarios in \ac{LOS} and \ac{NLOS} as well as \ac{COM} tasks. Therefore, to maximize the \ac{QoS}, the resource allocation is formulated as a constrained optimization problem whose objective function is the weighted sum of the utilities achieved with the assigned \ac{PAP} to each specific task. In this respect, the cumulative detection range is defined as a quality metric for search tasks, whereas for \ac{COM} tasks it is chosen as the range ensuring a desired channel capacity per bandwidth. Several case studies have been analyzed to prove the validity of the designed allocation strategy in challenging operational scenarios, ranging from the analysis of different priority weights selections to the study of the impact of the spatial selectivity of the antenna pointing angle. From the analyses of the results, the evidence is that the \ac{MPAR} tends to mostly allocate the available resources to the high priority tasks at the expense of the others. By doing so, it is ensured that the utilities for the most important tasks attain values close to their objectives, whereas for the remainder tasks a lower level of satisfaction is obtained.

Possible future researches could consider the extension of the framework to a multiface and/or multiband radar as well as to the multiradar systems. Moreover, the allocation of the beamformer weights to the different tasks is another valuable topic.

\section*{Acknowledgments}

The work of Augusto Aubry and Antonio De Maio was supported by the European Union under the Italian National Recovery and Resilience Plan (NRRP) of NextGenerationEU, partnership on ``Telecommunications of the Future'' (CUP J33C22002880001, PE00000001 - Program ``RESTART'').

\bibliographystyle{IEEEtran}
\bibliography{biblio}

\end{document}